\documentclass[prb,amsmath,amssymb,floatfix,twocolumn,citeautoscript]{revtex4}
\usepackage{bm}
\usepackage{epsfig}
\usepackage{subfigure}
\usepackage[usenames]{color}

\begin{document}

\title{Mechanism for nematic superconductivity in FeSe}

\author{Jian-Huang She$^1$, Michael J. Lawler$^{2, 1,3}$, and Eun-Ah Kim$^{1,3}$}

\affiliation{$^1$Department of Physics, Cornell University, Ithaca, NY 14853, USA \\
$^2$Department of physics, Binghamton University, Vestal, NY 13850, USA \\
$^3$Kavli Institute for Theoretical Physics, Kohn Hall
University Of California Santa Barbara CA 93106-4030, USA}

\begin{abstract}

Despite its seemingly simple composition and structure, the pairing mechanism of FeSe remains an open problem due to several striking phenomena. Among them are nematic order without  magnetic order, nodeless gap and unusual inelastic neutron spectra with a broad continuum, and gap anisotropy consistent with orbital selection of unknown origin. Here we propose a 
microscopic description of a nematic quantum spin liquid that reproduces key features of neutron spectra.
 We then study how the spin fluctuations of the local moments lead to pairing within a spin-fermion model. 
We find the resulting superconducting order parameter to be nodeless $s\pm d$-wave within each domain. Further
we show that orbital dependent Hund's coupling can readily capture observed gap anisotropy. Our prediction 
calls for inelastic neutron scattering in a detwinned sample.

\end{abstract}

\date{\today \ [file: \jobname]}

 \maketitle

The pairing mechanism and gap symmetry of 
bulk\cite{Paglione10, Hirschfeld11, Si16} and single layer\cite{QKXue12} FeSe is an open issue that 
inhibits an overarching understanding of iron-based superconductors. 
 Although a spin-fluctuation mediated pairing scenario is a broadly accepted mechanism in iron-based superconductors\cite{FWang11, Fernandes16}, much debate continues to focus around two distinct perspectives: weak coupling and strong coupling.
Weak coupling approaches 
 are sensitive to the band structure and generally predict dominantly $(\pi,0)$, $(0,\pi)$ spin density wave fluctuations that couple hole pockets to electron pockets in all Fe-pnictides as well as in bulk FeSe \cite{Chubukov16}. Strong coupling approaches take strong electron-electron correlations to generate quasi-localized moments that would interact with itinerant carriers. 
 
\begin{figure}[t]
\begin{centering}
\includegraphics[width=.35\textwidth]{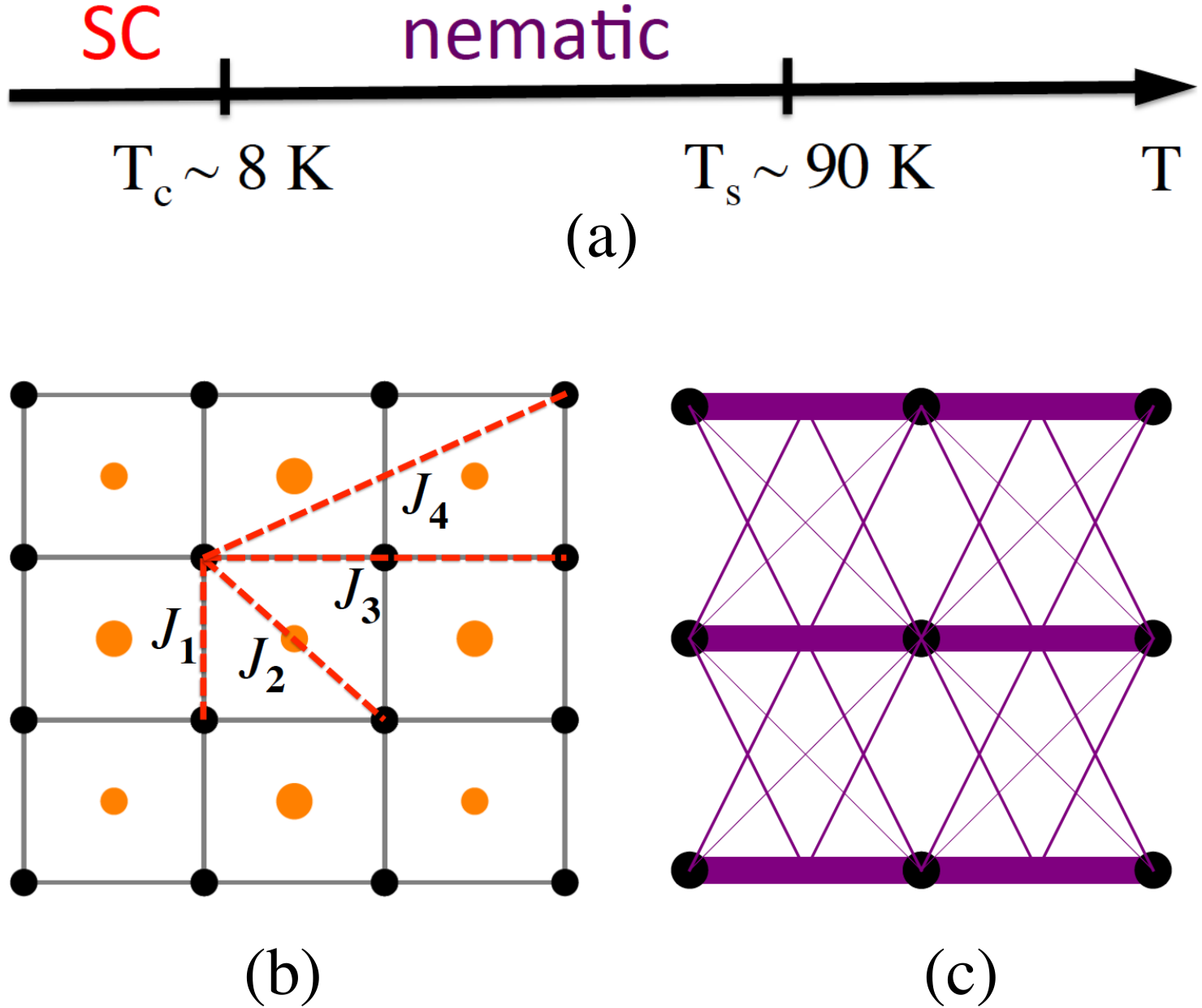} 
\end{centering}
\caption{(Color online) (a) Phase diagram of FeSe. (b) Lattice structure of FeSe. The black dots represent Fe atoms, and the orange dots represent Se atoms above and below the Fe plane. $J_{1-4}$ denote the exchange couplings.  (c) The nematic quantum spin liquid state of FeSe. The black dots represent spins, and the purple solid lines represent antiferromagnetic bonds, with their thickness proportional to bond strength. }
\label{Fig:model}
\end{figure}

 FeSe presents new challenges to both perspectives, including explaining its nematic order~\cite{McQueen09}(see Fig.~\ref{Fig:model}(a)), its absence of magnetism, its gapped but active spin fluctuations at $(\pi,\pi)$ in addition to $(\pi,0)$ \cite{Zhao16} and its nodeless superconducting gap. There have been much efforts to address these issues.  RPA based weak-coupling approaches focused on  implications of assumed nematic order~\cite{Hirschfeld15PRL, Hirschfeld15PRB}. Renormalization group approaches found the effective interactions promoting spin density wave to be also promoting orbital order \cite{Chubukov16, Xing16, Classen16}. Approaches focusing on sizable local moments~\cite{Gretarsson11} led to proposals of quadrupolar order accompanying nematic order \cite{Si15, Nevidomskyy16} and the proposal of 
a quasi-one dimensional quantum paramagnet state\cite{FWang15}  of AKLT (Affleck-Kennedy-Lieb-Tasaki)\cite{AKLT} type. Nevertheless, strikingly unique inelastic neutron spectra (INS) of FeSe evade the approaches so far one way or another.

The absence of the stripe order in FeSe has been attributed to the notion of frustration\cite{Mazin15,FWang15}.
Indeed FeSe is close to a classic situation for frustrated magnets in the much studied $J_1$-$J_2$ model\cite{Chandra90, Misguich04}(see Fig.~\ref{Fig:model}(b)). 
Interestingly, in systems that form stripe upon cooling, viewing the nematic state as thermally melted version of stripe was a very productive point of view \cite{Fernandes10}. Here we note that frustration from the competition between $J_1$ and $J_2$ has been long known to drive quantum melted versions of Neel and stripe orders giving rise to $C_4$ symmetric and $C_2$ symmetric (nematic) quantum spin liquids (QSL) respectively\cite{Haldane88,Read90}. 
Moreover DMRG studies on $J_1$-$J_2$ model noted an intermediate paramagnetic phase between stripe order and Neel order state \cite{Jiang09, Jiang12}. A recent DMRG study of $J_1$-$J_2$-$K_1$-$K_2$ spin model found a nematic quantum paramagnetic state between the Neel and stripe ordered states\cite{KYang16}. 
In this letter we propose a microscopic description of the frustration driven nematic quantum spin liquid (QSL)  state that amounts to quantum melted stripe and
captures the observed INS. We then investigate the implication of dramatically anisotropic spin-fluctuation spectra of the proposed state on the nature of superconductivity.

\begin{figure}[h]
\begin{centering}
\includegraphics[width=.5\textwidth]{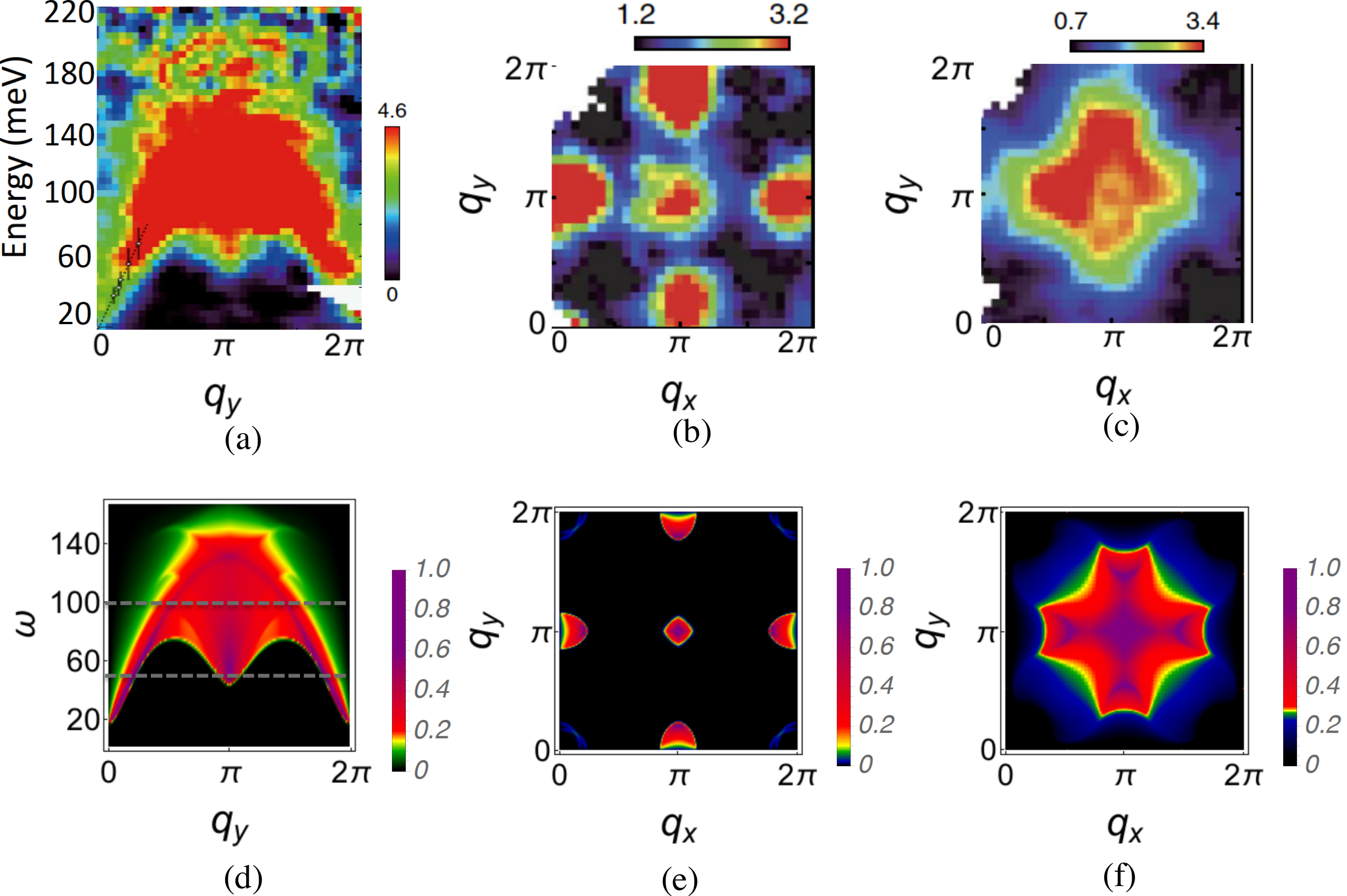}
\end{centering}
\caption{(Color online) (a, b, c): Neutron scattering results for the dynamic spin structure factor ${\cal S}(q_x, q_y, \omega)$ at $q_x=\pi$ (a), $\omega=50, 100$ meV (b, c) \cite{Zhao16}. (d, e, f): The corresponding results from theoretical calculation using SBMFT summed over two nematic domains.}
\label{Fig:chi}
\end{figure}

In FeSe, there is evidence that local moments~\cite{Gretarsson11} coexist with itinerant carriers of all three $t_{2g}$ orbitals\cite{Nakayama14, Watson15, Suzuki15}.
In order to capture the dual character \cite{Moon10} 
we turn to a spin-fermion model~\cite{Weng09, Phillips10, Ku10, Littlewood13, Dagotto14, DHLee15}: ${\cal H}={\cal H}_c+{\cal H}_S+{\cal H}_{\rm int}$, where ${\cal H}_c$ and ${\cal H}_S$ describe the itinerant carriers and local moments respectively that are coupled through  ${\cal H}_{\rm int}$. 
 For the spin model 
\begin{equation}
{\cal H}_S= \sum_{ij} J_{ij}{\bm S}_i\cdot{\bm S}_j,
\end{equation}
with exchange interactions $J_{ij}$ on a square lattice (Fig.1b), the two dominant interactions 
are the nearest-neighbor $J_1$ and the next-nearest-neighbor $J_2$ exchange interactions as in other Fe-based superconductors \cite{Chen08, Xu08}. But due to the near itinerancy of the core electrons, longer range terms  
are also expected~\cite{Mazin15}.

The $J_1$-$J_2$ model has been extensively studied both classically and quantum mechanically
(see Refs.[\onlinecite{Chandra90, Misguich04, Jiang09, Jiang12}]). Within classical models the role of frustration is clear from the fact that the model can be recast as
${\cal H}_S=J_2\sum ({\bm S}_1+{\bm S}_2+{\bm S}_3+{\bm S}_4)^2$ up to a constant at $J_2=J_1/2$ point, where ${\bm S}_{1-4}$ are the four spins on each plaquette $\langle 1234\rangle$ and the summation is over all plaquettes.
Classical ground state with vanishing total spin on each plaquette property leads to a zero mode at each wave vector on the Brillouin zone boundary\cite{Misguich04} and so the model is highly frustrated.
With quantum effects of small spin $S$ the frustration effects are not limited to the fine tuned point of $J_2=J_1/2$. Unfortunately, a controlled theoretical study for quantum spins for such frustrated spin systems is challenging. Hence we will restrict ourselves to mean field theories and choose an ansatz that (1) agrees with the observed inelastic neutron spectrum \cite{Zhao16}, and (2) the ordering tendencies obey the classical condition of ${\bm S}_1+{\bm S}_2+{\bm S}_3+{\bm S}_4=0$ on a plaquette.

 A prominent feature of the INS data \cite{Zhao16} is its broad and gapped continuum of spectral weight (Fig.\ref{Fig:chi}a) without any one-magnon branch. Intriguingly such a continuum is expected in a QSL with deconfined spinons in two-dimension in an insulating magnetic system \cite{Balents10}. Indeed it is a common feature of slave-particle mean field theories. So we will choose Schwinger boson mean field theory (SBMFT)~\cite{Arovas88} as our mean field theory approach. Additional features of Fig.~\ref{Fig:chi} (a-c) we aim to capture include:
\begin{itemize}
\item The simultaneous presence of both $(\pi, \pi)$ spin fluctuations and $(\pi, 0)$, $(0, \pi)$ spin fluctuations. 
\item The quasi-one-dimensional dispersion $\omega\sim \sin k_y$  \cite{Muller81, Lake13, Caux14} found in the shape of the upper and lower bounds.
\item The observed cross-shaped spectrum around $(\pi, \pi)$.
\end{itemize}

\begin{figure}[h]
\begin{centering}
\includegraphics[width=.4\textwidth]{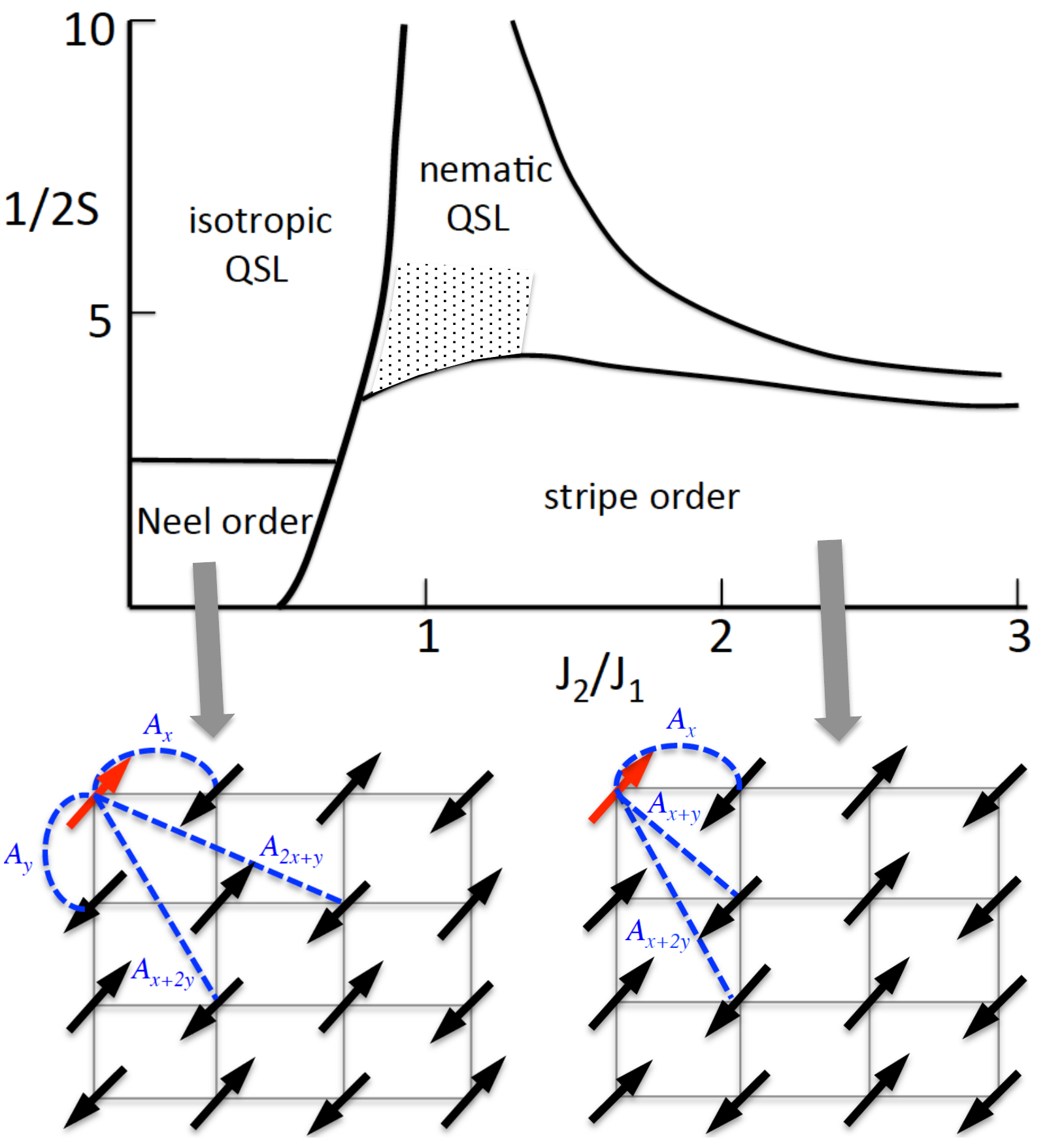}
\end{centering}
\caption{(Color online) (top) The SBMFT phase diagram of the $J_1$-$J_2$ model computed using Sp($N$) Schwinger bosons \cite{Sachdev91}. Our mean field ansatz corresponds to the 
shaded region. (bottom) The spin configurations in the two long-range ordered phases.  The blue dashed lines represent the mean field bonds $A_{{\bf r},{\bf r'}} = \langle b^\dagger_{{\bf r}\uparrow}b^\dagger_{{\bf r'}\downarrow}-b^\dagger_{{\bf r}\downarrow}b^\dagger_{{\bf r'}\uparrow}\rangle$, connecting a spin (red arrow) with its neighboring spins (black arrows). Here Neel order, stripe order, isotropic QSL, nematic QSL correspond respectively to $(\pi, \pi)_{\rm LRO}$, $(\pi, 0)_{\rm LRO}$, $(\pi, \pi)_{\rm SRO}$, $(\pi, 0)_{\rm SRO}$ in \cite{Sachdev91}.  We choose the ordinary Sp(1) = SU(2) spins since the distinction with general $N$ is unimportant for our purposes.}
\label{Fig:spin}
\end{figure}

To find these features in a SBMFT, we turn to the known \cite{Sachdev91} SBMFT phase diagram of the $J_1$-$J_2$ model (Fig. \ref{Fig:spin}). 
Note that the Neel and stripe long range order for small $J_2/J_1$ and large $J_2/J_1$ are expected\cite{Read90} to melt into $C_4$ symmetric and $C_2$ symmetric QSL's respectively (see Fig.~\ref{Fig:spin}).
Hence the shaded region near the phase boundary between $C_4$ symmetric QSL, $C_2$ symmetric QSL and the stripe ordered phase will capture all of the above features. Specifically, states in this region will support a dynamic spin structure factor with 1d-like dispersion and cross-shaped spectrum 
\emph{assuming twin domains of the stripe state are averaged over in the INS data}. To account for the itinerancy of the electrons, we extend  
an ansatz within the shaded region of Fig. \ref{Fig:spin}  
with additional neighbor couplings.

To construct the ansatz, we now turn briefly to the specifics of SBMFT. In Schwinger boson representation, each spin ${\bm S}_{\bf r}$  is represented by two bosonic operators $b_{{\bf r}\sigma}$, $\sigma=\uparrow,\downarrow$ with the constraint $\sum_\sigma b^\dagger_{{\bf r}\sigma}b_{{\bf r}\sigma}=2S$. The spin operator is then ${\bm S}_{\bf r}=\frac{1}{2}\sum_{\sigma\sigma'}b^\dagger_{{\bf r}\sigma}{\bm \sigma}_{\sigma\sigma'}b_{{\bf r}\sigma'}$, with $\bm \sigma$ the Pauli matrices. We can then expand $H_{{\bf r},{\bf r'}} \equiv J_{{\bf r},{\bf r'}}{\bm S}_{\bf r}\cdot{\bm S}_{\bf r'}$ in terms of the spin singlet operator $A^\dagger_{{\bf r},{\bf r'}}=b^\dagger_{{\bf r}\uparrow}b^\dagger_{{\bf r'}\downarrow}-b^\dagger_{{\bf r}\downarrow}b^\dagger_{{\bf r'}\uparrow}$ to obtain $H_{{\bf r},{\bf r'}} = -J_{{\bf r},{\bf r'}}\frac{1}{2}A^\dagger_{{\bf r},{\bf r'}}A_{{\bf r},{\bf r'}}+S^2$. Finally, 
we mean-field decompose $H_{{\bf r},{\bf r'}}$ and introduce mean fields $\langle A_{{\bm r},{\bm r'}}\rangle$ using $A^\dagger_{{\bf r},{\bf r'}}A_{{\bf r},{\bf r'}} = \langle A^\dagger_{{\bf r},{\bf r'}}\rangle A_{{\bf r},{\bf r'}} + A^\dagger_{{\bf r},{\bf r'}}\langle A_{{\bf r},{\bf r'}}\rangle - \langle A^\dagger_{{\bf r},{\bf r'}}\rangle\langle A_{{\bf r},{\bf r'}}\rangle$. We will further assume the bosons do not
 condense for we are interested in the spin liquid phase.

Defining $A_{\hat{\mu}} \equiv \langle A_{{\bm r},{\bm r}+{\hat \mu}}\rangle$, we keep  $A_{\hat x}\neq 0$ and the diagonals $A_{\hat x \pm \hat y}\neq 0$ and $A_{{\hat x}\pm 2{\hat y}}\neq 0$
for states in the shaded region of Fig.~\ref{Fig:spin}. 
The fourth neighbor term can be understood as a result of the competition between Neel and stripe states: it is a bond that is favored by both the $(\pi,\pi)$ N\'eel state and the $(\pi,0)$/$(0,\pi)$ stripe state.
The result is a state with the same projective symmetry group as the Read and Sachdev state used in the phase diagram of Fig. \ref{Fig:spin}. It is a ``zero flux state" in that the smallest loop has zero ``flux" obeying the so-called flux expulsion principle \cite{Tchernyshyov06} and hence energetically competitive. 
Most importantly it is a state in which translational symmetry is restored by quantum melting stripe into $C_2$ symmetric nematic QSL state.

We can then calculate 
the dynamic spin structure factor ${\cal S}_{{\bm q}\omega}\equiv{\rm Im}\langle S^z({\bm q}, \omega) S^z(-{\bm q}, \omega)\rangle$ associated with our ansatz.
 At $T=0$, it is of the form \cite{Auerbach88}
\begin{equation}
{\cal S}_{{\bm q}, \omega} \sim \sum_{\bm k}\left\{ \cosh\left[ 2\left(\theta_{\bm k}+\theta_{{\bm k}+{\tilde{\bm q}}} \right)\right]-1\right\}\delta\left( \omega_{\bm k}+\omega_{{\bm k}+{\tilde{\bm q}}}-|\omega|\right),
\label{Eq:SSF}
\end{equation}
where $\theta_{\bm k}$ is the angle in the Bogoliubov transformation of SBMFT (see SM1 for explicit expression), and ${\tilde{\bm q}}={\bm q}-(\pi, 0)$ arises because of a standard unitary transformation we carried out on the B sublattice for simplicity. The results summing over two domains are plotted in Fig.~\ref{Fig:chi}(d-f). They capture the basic features of the neutron spectra: (1) The spectrum is gapped (Fig.~\ref{Fig:chi}d), as a result of the absence of long range magnetic ordering. 
 (2) Both $(\pi, \pi)$ and $(\pi, 0)$/$(0, \pi)$ spin fluctuations are present (Fig.~\ref{Fig:chi}d, e).   
  (3) The spectrum displays the novel feature of continuum with the bounds exhibiting quasi-one-dimensional dispersion (Fig.~\ref{Fig:chi}d).
  
 A sharp prediction of our model is the dramatic suppression of 
 spectral weight around $(0, q_y)$ in a detwinned sample ($(q_x, 0)$ for the other domain). This means at low energies there are weights at say $(\pi, \pi)$ and $(\pi, 0)$, but not at $(0, \pi)$. 
 By contrast, in an orbital order driven picture for nematic ordering, there is only a weak anisotropy in the spin-structure factor with the spectral weight at $(\pi, \pi)$, $(0, \pi)$ and $(\pi, 0)$ of roughly the same magnitude even in a single nematic domain\cite{Hirschfeld15PRL,Hirschfeld15PRB}.
 Such a distinction has profound implications for pairing.  When the degree of anisotropy in the momentum distribution of the spin spectra is mild, pairing interactions with different ${\bm q}$-wavevectors compete, leading to nodes \cite{Hirschfeld15PRL,Hirschfeld15PRB}. On the other hand, the strong anisotropy in the spectral weight distribution in our SBMFT ansatz quenches such competition removing any need for a superconducting gap node.

\begin{figure}[h]
\begin{centering}
\includegraphics[width=.4\textwidth]{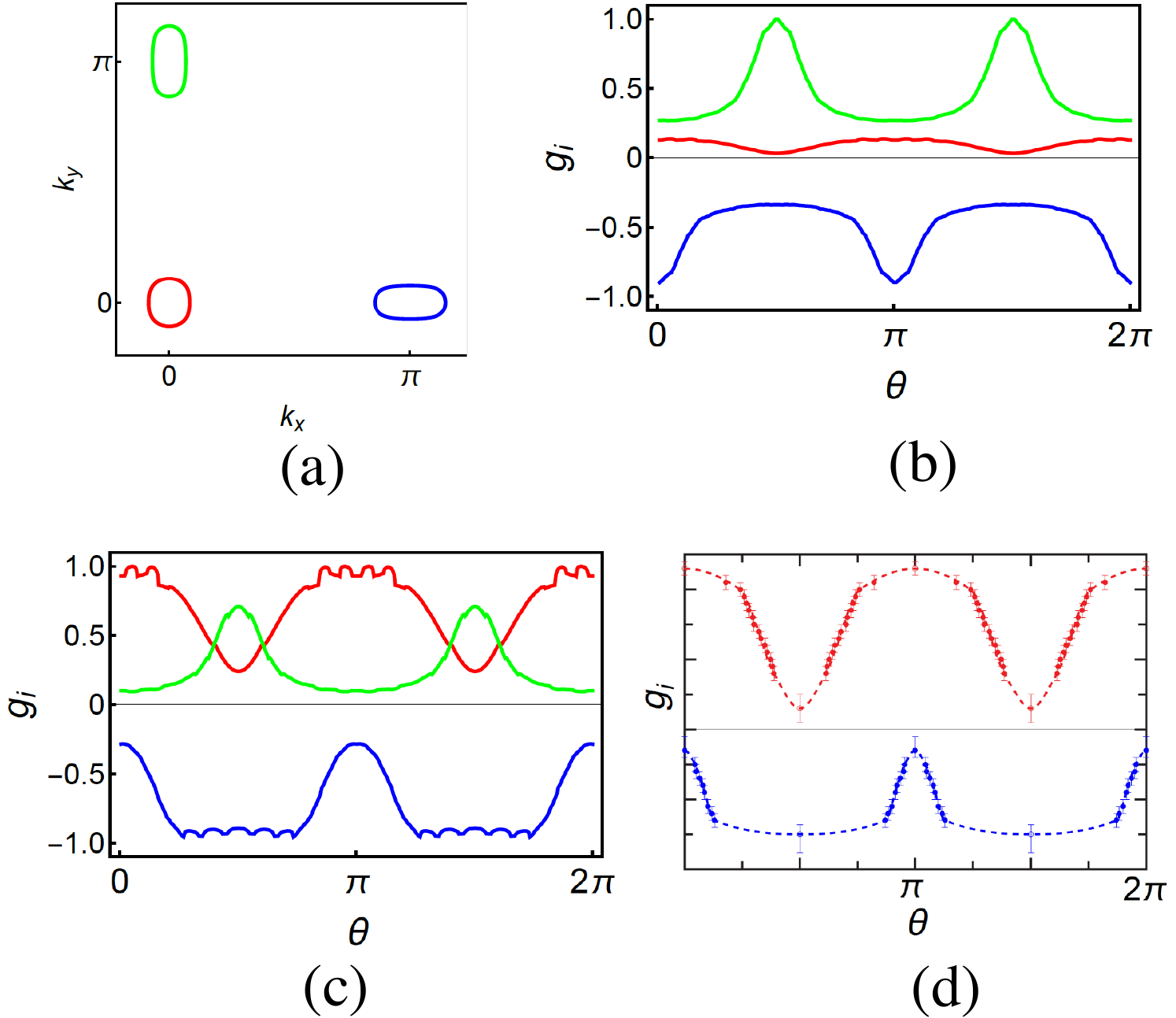}
\end{centering}
\caption{(Color online) (a): The Fermi surface. (b, c): The gap symmetry function on different Fermi pockets for three-band models with  $J_{yz}=J_{zx}=J_{xy}=1$ (b) and  $J_{yz}=J_{zx}=1$, $J_{xy}=0.4$ (c). (d): The gap function observed in the recent STM measurements \cite{Sprau16}. }
\label{Fig:gap-structure}
\end{figure}

We now turn to the itinerant degrees of freedom to study nematicity and superconductivity. Their kinetic energy is given by a tight-binding model:
 \begin{equation}
{\cal H}_c=\sum_{{\bm k},\alpha\beta,\nu}\epsilon_{\alpha\beta}^{\mu\nu}({\bm k})c^\dagger_{\alpha\mu}({\bm k})c_{\beta\nu}({\bm k}),
\end{equation}
where $c^\dagger_{\alpha\mu}({\bm k})$ creates an itinerant electron with momentum ${\bm k}$, spin $\mu$ and orbital index $\alpha$. 
The Fermi surface of FeSe consists of two electron pockets around the M points and one hole pocket around the $\Gamma$ point \cite{Nakayama14, Watson15, Suzuki15}. Following \cite{Vafek13, Fernandes16}, we take a simple symmetry based approach of expanding the dispersion $\epsilon_{\alpha\beta}^{\mu\nu}({\bm k})$ around the Fermi surface. It is known experimentally that the spectral weight of the low energy states are predominantly from $d_{yz}$ and $d_{zx}$ around the $\Gamma$ point, from $d_{yz}$ and $d_{xy}$ around $(\pi, 0)$, from $d_{zx}$ and $d_{xy}$ around $(0, \pi)$. We consider the corresponding intra- and inter-orbital hopping terms. Furthermore we include on-site nematicity and spin-orbit coupling to produce the band splitting that gives rise to a single hole pocket around $\Gamma$. The resulting simplified Fermi surface is shown in Fig.\ref{Fig:gap-structure}a, see SM2 for explicit parameters. \footnote{Our rather simple band structure allows largely analytic calculation at the expense of missing out quantiative details such as large mismatch in the pocket sizes as found in quantum oscillation (see SM4).}

The itinerant electrons couple to the local moments via the ferromagnetic Hund's coupling \cite{Phillips10}:
\begin{equation}
{\cal H}_{\rm int}=-\sum_{i,\alpha,\mu\nu}J_{\alpha}{\bm S}_i\cdot c^\dagger_{i\alpha\mu}{\bm \sigma}_{\mu\nu}c_{i\alpha\nu},
\end{equation}
where ${\bm \sigma}$ represents the vector of Pauli matrices, and $J_\alpha>0$ denote the Hund's couplings. 
Since the Hund's couplings depend on the overlap of the itinerant electron wave function with the local moment, they are generally different for different orbitals. 
\footnote{Coupling to itinerant electrons generates a self-energy for the local moment propagator, giving rise to Landau damping (see SM3). However the strength of the Hund's couplings can be estimated to be much weaker than the spin exchange interaction (see SM3). Hence coupling to itinerant electrons will not significantly modify the local moment spin susceptibility.}

 Note that the proposed nematic QSL state induces nematicity in the charge sector. 
  For instance non-zero 
  $\langle A_{{\bm r},{\bm r}\pm {\hat x}}\rangle$ in the nematic QSL state generates 
  an interaction among conduction electrons along the $x$-direction,  
  which drives bond-centered nematic order with 
  $\varphi_c\equiv  \langle c^\dagger_{{\bm r}+{\hat x},\alpha}c_{{\bm r},\alpha} -c^\dagger_{{\bm r}+{\hat y},\alpha}c_{{\bm r},\alpha} \rangle \neq 0$ below the temperature at which the nematic QSL develops. The observed nematic transition at $T_s\sim 90K$ \cite{McQueen09} is consistent with this picture. Furthermore, $\varphi_c$ linearly couples to 
   $\varphi_o\equiv \frac{n_{zx}-n_{yz}}{n_{zx}+n_{yz}}$, where $n_{zx, yz}$ denote occupation of $zx$ and $yz$ orbitals, and $\varphi_s \equiv M_x^2-M_y^2$, where ${\bm M}$ represents the magnetic moment. These different measures of nematicity are consistent with orbital imbalance 
   observed in ARPES \cite{Nakayama14, Watson15, Suzuki15} ($\varphi_o\neq 0$) and the  observed NMR resonance line splitting \cite{Brink15} ($\varphi_s\neq 0$).

Furthermore, the nematic spin fluctuations in the proposed QSL state mediate pairing 
and the resulting gap structure can be determined via standard a mean field procedure (see SM4). An immediate observation is that  non-universal aspects of the gap structure such as relative gap strength of each pocket and the $T_c$ are sensitive to strength of the Hund's couplings $J$'s (see Fig.~\ref{Fig:gap-structure}b,c). Nevertheless the gap functions resulting from our model share the following generic features: (1) The gap is generically nodeless as a result of severe anisotropy of the spin fluctuations in the nematic QSL state. In particular, the near absence of spin fluctuations around say $(0, \pi)$ for one nematic domain renders the determination of gap sign in different pockets unfrustrated. In contrast, in the itinerant model where $(\pi, \pi)$, $(\pi, 0)$ and $(0, \pi)$ spin fluctuations are close in magnitude they compete for deciding the sign structure of the gap causing nodal gap structures. (2) The gap is deeply anisotropic due to the 
variation of orbital content around each Fermi pocket. The resulting nodeless but very anisotropic gap structure explains the seemingly contradictory experimental results of STM \cite{QKXue11, Matsuda14}, penetration depth \cite{Matsuda14}, thermal conductivity measurements \cite{Taillefer16}, observing low energy excitations \cite{QKXue11, Matsuda14} despite the evidence of a full gap \cite{Taillefer16, Sprau16}. (3) 
The gap changes sign from pocket to pocket. This is consistent (see SM4) with the observation of sharp spin resonance in the superconducting state \cite{JZhao15}. 
More specifically, our gap function is a combination of $d$-wave as induced by $(\pi, \pi)$ spin fluctuations and $s_{\pm}$ as induced by $(\pi, 0)$ spin fluctuations.
 We consider a single nematic domain, where the pairing interaction concentrates around $(\pi, q_y)$. Two examples of the gap function (in arbitrary units) are shown in Fig.\ref{Fig:gap-structure}b,c, where $J_{xy}=J_{yz}=J_{zx}$ and $J_{xy}=0.4 J_{yz}=0.4 J_{zx}$ respectively.

Now we turn to the question of orbital dependence of the Hund's coupling. Fig.~\ref{Fig:gap-structure}b,c shows that the orbital dependent Hund's coupling can alter the relative magnitude and anisotropy of gap functions at different Fermi pockets (while the gap is predominantly $d$-wave in Fig.~\ref{Fig:gap-structure}b, $d$- and $s$-wave are at par in Fig.~\ref{Fig:gap-structure}c). 
Since the Hund's coupling requires overlap of the wave-function between the conduction electrons and local moments, significantly lower spectral weight of $d_{xy}$ orbitals \cite{DHLu15} implies $J_{xy}\ll J_{zx}, J_{yz}$.

Indeed, the gap function with such orbital dependent Hund's coupling shows remarkable resemblance to the gap structure observed by recent STM measurements\cite{Sprau16} (see Fig.~\ref{Fig:gap-structure}c,d).  In Ref.~ \textcite{Sprau16} the observed pocket specific gap anisotropy was interpreted as resulting from orbital-selective pairing of unknown microscopic origin.
In our model, such orbital selective pairing arise from orbital dependence in the Hund's coupling $J_{xy}<J_{yz}=J_{zx}$, reflecting much smaller weight of $d_{xy}$ orbitals in the conduction electrons~\cite{DHLu15}. This orbital dependent Hund's coupling amplify the role of $(\pi,0)$ spin fluctuation in pairing despite larger spectral weight at $(\pi,\pi)$, which is consistent with the observation of sharp spin resonance at $(\pi,0)$~\cite{JZhao15} (see SM4 for further discussion).

In conclusion, we propose a nematic QSL state description of FeSe that explains 
the basic phenomenology of FeSe: (1) Spin dynamics observed in Ref.\onlinecite{Zhao16} assuming it is averaged over domains, (2) nematic transition without mangetic ordering, (3) highly anisotropic fully gapped superconducting gap. 
The central assumption that neutron scattering is averaging over domains could be tested in a detwinned neutron experiment. Orbital dependent Hund's coupling mechanisms for orbital selective pairing in bulk FeSe further offers new insight regarding higher $T_c$ observed in 
mono-layer FeSe and K-doped FeSe. As we show in SM4, larger $J_{xy}$ that enables conduction electrons to utilize $(\pi,\pi)$ spin fluctuation with larger intensity and higher characteristic frequency leads to higher transition temperature (as high as ~47K). Combined with the observation that spectral weight of the $d_{xy}$ orbitals in the conduction electrons is much higher in the higher $T_c$ settings of  mono-layer FeSe and K-doped FeSe~\cite{DHLu15}, it is conceivable these systems make better use of already more prominent $(\pi,\pi)$ fluctuation to achieve higher $T_c$. We note here that the nematic QSL state we propose is distinct from the proposal of Ref. \onlinecite{FWang15} in that it contains no one-magnon branch of excitations (see SM5) although both proposals start from strong coupling perspective and spin ground states lacking any form of magnetic order.
Finally, although we used SBMFT as a calculational crutch to capture the spin wave continuum, the ultimate fate of spinons in this spin system coupled to itinerant electrons needs further study. Interestingly, such a state with spinons coexisting with conduction electrons would resemble the FL* state first proposed in Refs.~\onlinecite{Senthil03, Senthil04} that has recently been revisited using DMRG \cite{Sachdev16}.

\vspace{5mm}
\noindent {\bf Acknowledgements} We thank Andrey Chubukov, J.C. Davis, Rafael Fernandez, Yong Baek Kim, Steve Kivelson, Igor Mazin, Andriy Nevidomskyy, Subir Sachdev, Doug Scalapino, Qimiao Si, Fa Wang   for discussions. E-AK  and J-HS were  supported by the U.S. Department of Energy, Office of Basic Energy Sciences,  Division of Materials Science and Engineering  under  Award  DE-SC0010313.   E-AK  also  acknowledges Simons Fellow in Theoretical Physics Award\#392182. E-AK and MJL acknowledge hospitality of the KITP supported by Grant No. NSF PHY11-25915.

\bibliographystyle{apsrev}
\bibliography{strings,refs}

\begin{thebibliography}{70}
\expandafter\ifx\csname natexlab\endcsname\relax\def\natexlab#1{#1}\fi
\expandafter\ifx\csname bibnamefont\endcsname\relax
  \def\bibnamefont#1{#1}\fi
\expandafter\ifx\csname bibfnamefont\endcsname\relax
  \def\bibfnamefont#1{#1}\fi
\expandafter\ifx\csname citenamefont\endcsname\relax
  \def\citenamefont#1{#1}\fi
\expandafter\ifx\csname url\endcsname\relax
  \def\url#1{\texttt{#1}}\fi
\expandafter\ifx\csname urlprefix\endcsname\relax\def\urlprefix{URL }\fi
\providecommand{\bibinfo}[2]{#2}
\providecommand{\eprint}[2][]{\url{#2}}

\bibitem[{\citenamefont{{Paglione} and {Greene}}(2010)}]{Paglione10}
\bibinfo{author}{\bibfnamefont{J.}~\bibnamefont{{Paglione}}} \bibnamefont{and}
  \bibinfo{author}{\bibfnamefont{R.~L.} \bibnamefont{{Greene}}},
  \bibinfo{journal}{Nature Physics} \textbf{\bibinfo{volume}{6}},
  \bibinfo{pages}{645} (\bibinfo{year}{2010}), \eprint{1006.4618}.

\bibitem[{\citenamefont{Hirschfeld et~al.}(2011)\citenamefont{Hirschfeld,
  Korshunov, and Mazin}}]{Hirschfeld11}
\bibinfo{author}{\bibfnamefont{P.~J.} \bibnamefont{Hirschfeld}},
  \bibinfo{author}{\bibfnamefont{M.~M.} \bibnamefont{Korshunov}},
  \bibnamefont{and} \bibinfo{author}{\bibfnamefont{I.~I.} \bibnamefont{Mazin}},
  \bibinfo{journal}{Reports on Progress in Physics}
  \textbf{\bibinfo{volume}{74}}, \bibinfo{pages}{124508}
  (\bibinfo{year}{2011}),
  \urlprefix\url{http://stacks.iop.org/0034-4885/74/i=12/a=124508}.

\bibitem[{\citenamefont{{Si} et~al.}(2016)\citenamefont{{Si}, {Yu}, and
  {Abrahams}}}]{Si16}
\bibinfo{author}{\bibfnamefont{Q.}~\bibnamefont{{Si}}},
  \bibinfo{author}{\bibfnamefont{R.}~\bibnamefont{{Yu}}}, \bibnamefont{and}
  \bibinfo{author}{\bibfnamefont{E.}~\bibnamefont{{Abrahams}}},
  \bibinfo{journal}{Nature Reviews Materials} \textbf{\bibinfo{volume}{1}},
  \bibinfo{pages}{16017} (\bibinfo{year}{2016}), \eprint{1604.03566}.

\bibitem[{\citenamefont{{Wang} et~al.}(2012)\citenamefont{{Wang}, {Li},
  {Zhang}, {Zhang}, {Zhang}, {Li}, {Ding}, {Ou}, {Deng}, {Chang}
  et~al.}}]{QKXue12}
\bibinfo{author}{\bibfnamefont{Q.-Y.} \bibnamefont{{Wang}}},
  \bibinfo{author}{\bibfnamefont{Z.}~\bibnamefont{{Li}}},
  \bibinfo{author}{\bibfnamefont{W.-H.} \bibnamefont{{Zhang}}},
  \bibinfo{author}{\bibfnamefont{Z.-C.} \bibnamefont{{Zhang}}},
  \bibinfo{author}{\bibfnamefont{J.-S.} \bibnamefont{{Zhang}}},
  \bibinfo{author}{\bibfnamefont{W.}~\bibnamefont{{Li}}},
  \bibinfo{author}{\bibfnamefont{H.}~\bibnamefont{{Ding}}},
  \bibinfo{author}{\bibfnamefont{Y.-B.} \bibnamefont{{Ou}}},
  \bibinfo{author}{\bibfnamefont{P.}~\bibnamefont{{Deng}}},
  \bibinfo{author}{\bibfnamefont{K.}~\bibnamefont{{Chang}}},
  \bibnamefont{et~al.}, \bibinfo{journal}{Chinese Physics Letters}
  \textbf{\bibinfo{volume}{29}}, \bibinfo{eid}{037402} (\bibinfo{year}{2012}),
  \eprint{1201.5694}.

\bibitem[{\citenamefont{Wang and Lee}(2011)}]{FWang11}
\bibinfo{author}{\bibfnamefont{F.}~\bibnamefont{Wang}} \bibnamefont{and}
  \bibinfo{author}{\bibfnamefont{D.-H.} \bibnamefont{Lee}},
  \bibinfo{journal}{Science} \textbf{\bibinfo{volume}{332}},
  \bibinfo{pages}{200} (\bibinfo{year}{2011}), ISSN \bibinfo{issn}{0036-8075},
  \eprint{http://science.sciencemag.org/content/332/6026/200.full.pdf},
  \urlprefix\url{http://science.sciencemag.org/content/332/6026/200}.

\bibitem[{\citenamefont{{Fernandes} and {Chubukov}}(2016)}]{Fernandes16}
\bibinfo{author}{\bibfnamefont{R.~M.} \bibnamefont{{Fernandes}}}
  \bibnamefont{and} \bibinfo{author}{\bibfnamefont{A.~V.}
  \bibnamefont{{Chubukov}}}, \bibinfo{journal}{ArXiv e-prints}
  (\bibinfo{year}{2016}), \eprint{1607.00865}.

\bibitem[{\citenamefont{Chubukov et~al.}(2016)\citenamefont{Chubukov, Khodas,
  and Fernandes}}]{Chubukov16}
\bibinfo{author}{\bibfnamefont{A.~V.} \bibnamefont{Chubukov}},
  \bibinfo{author}{\bibfnamefont{M.}~\bibnamefont{Khodas}}, \bibnamefont{and}
  \bibinfo{author}{\bibfnamefont{R.~M.} \bibnamefont{Fernandes}},
  \bibinfo{journal}{Phys. Rev. X} \textbf{\bibinfo{volume}{6}},
  \bibinfo{pages}{041045} (\bibinfo{year}{2016}),
  \urlprefix\url{http://link.aps.org/doi/10.1103/PhysRevX.6.041045}.

\bibitem[{\citenamefont{McQueen et~al.}(2009)\citenamefont{McQueen, Williams,
  Stephens, Tao, Zhu, Ksenofontov, Casper, Felser, and Cava}}]{McQueen09}
\bibinfo{author}{\bibfnamefont{T.~M.} \bibnamefont{McQueen}},
  \bibinfo{author}{\bibfnamefont{A.~J.} \bibnamefont{Williams}},
  \bibinfo{author}{\bibfnamefont{P.~W.} \bibnamefont{Stephens}},
  \bibinfo{author}{\bibfnamefont{J.}~\bibnamefont{Tao}},
  \bibinfo{author}{\bibfnamefont{Y.}~\bibnamefont{Zhu}},
  \bibinfo{author}{\bibfnamefont{V.}~\bibnamefont{Ksenofontov}},
  \bibinfo{author}{\bibfnamefont{F.}~\bibnamefont{Casper}},
  \bibinfo{author}{\bibfnamefont{C.}~\bibnamefont{Felser}}, \bibnamefont{and}
  \bibinfo{author}{\bibfnamefont{R.~J.} \bibnamefont{Cava}},
  \bibinfo{journal}{Phys. Rev. Lett.} \textbf{\bibinfo{volume}{103}},
  \bibinfo{pages}{057002} (\bibinfo{year}{2009}),
  \urlprefix\url{http://link.aps.org/doi/10.1103/PhysRevLett.103.057002}.

\bibitem[{\citenamefont{Wang et~al.}(2016{\natexlab{a}})\citenamefont{Wang,
  Shen, Pan, Zhang, Ikeuchi, Iida, Christianson, Walker, Adroja, Abdel-Hafiez
  et~al.}}]{Zhao16}
\bibinfo{author}{\bibfnamefont{Q.}~\bibnamefont{Wang}},
  \bibinfo{author}{\bibfnamefont{Y.}~\bibnamefont{Shen}},
  \bibinfo{author}{\bibfnamefont{B.}~\bibnamefont{Pan}},
  \bibinfo{author}{\bibfnamefont{X.}~\bibnamefont{Zhang}},
  \bibinfo{author}{\bibfnamefont{K.}~\bibnamefont{Ikeuchi}},
  \bibinfo{author}{\bibfnamefont{K.}~\bibnamefont{Iida}},
  \bibinfo{author}{\bibfnamefont{A.~D.} \bibnamefont{Christianson}},
  \bibinfo{author}{\bibfnamefont{H.~C.} \bibnamefont{Walker}},
  \bibinfo{author}{\bibfnamefont{D.~T.} \bibnamefont{Adroja}},
  \bibinfo{author}{\bibfnamefont{M.}~\bibnamefont{Abdel-Hafiez}},
  \bibnamefont{et~al.}, \bibinfo{journal}{Nat Commun}
  \textbf{\bibinfo{volume}{7}} (\bibinfo{year}{2016}{\natexlab{a}}),
  \urlprefix\url{http://dx.doi.org/10.1038/ncomms12182}.

\bibitem[{\citenamefont{Mukherjee et~al.}(2015)\citenamefont{Mukherjee,
  Kreisel, Hirschfeld, and Andersen}}]{Hirschfeld15PRL}
\bibinfo{author}{\bibfnamefont{S.}~\bibnamefont{Mukherjee}},
  \bibinfo{author}{\bibfnamefont{A.}~\bibnamefont{Kreisel}},
  \bibinfo{author}{\bibfnamefont{P.~J.} \bibnamefont{Hirschfeld}},
  \bibnamefont{and} \bibinfo{author}{\bibfnamefont{B.~M.}
  \bibnamefont{Andersen}}, \bibinfo{journal}{Phys. Rev. Lett.}
  \textbf{\bibinfo{volume}{115}}, \bibinfo{pages}{026402}
  (\bibinfo{year}{2015}),
  \urlprefix\url{http://link.aps.org/doi/10.1103/PhysRevLett.115.026402}.

\bibitem[{\citenamefont{Kreisel et~al.}(2015)\citenamefont{Kreisel, Mukherjee,
  Hirschfeld, and Andersen}}]{Hirschfeld15PRB}
\bibinfo{author}{\bibfnamefont{A.}~\bibnamefont{Kreisel}},
  \bibinfo{author}{\bibfnamefont{S.}~\bibnamefont{Mukherjee}},
  \bibinfo{author}{\bibfnamefont{P.~J.} \bibnamefont{Hirschfeld}},
  \bibnamefont{and} \bibinfo{author}{\bibfnamefont{B.~M.}
  \bibnamefont{Andersen}}, \bibinfo{journal}{Phys. Rev. B}
  \textbf{\bibinfo{volume}{92}}, \bibinfo{pages}{224515}
  (\bibinfo{year}{2015}),
  \urlprefix\url{http://link.aps.org/doi/10.1103/PhysRevB.92.224515}.

\bibitem[{\citenamefont{{Xing} et~al.}(2016)\citenamefont{{Xing}, {Classen},
  {Khodas}, and {Chubukov}}}]{Xing16}
\bibinfo{author}{\bibfnamefont{R.-Q.} \bibnamefont{{Xing}}},
  \bibinfo{author}{\bibfnamefont{L.}~\bibnamefont{{Classen}}},
  \bibinfo{author}{\bibfnamefont{M.}~\bibnamefont{{Khodas}}}, \bibnamefont{and}
  \bibinfo{author}{\bibfnamefont{A.~V.} \bibnamefont{{Chubukov}}},
  \bibinfo{journal}{ArXiv e-prints}  (\bibinfo{year}{2016}),
  \eprint{1611.03912}.

\bibitem[{\citenamefont{{Classen} et~al.}(2016)\citenamefont{{Classen}, {Xing},
  {Khodas}, and {Chubukov}}}]{Classen16}
\bibinfo{author}{\bibfnamefont{L.}~\bibnamefont{{Classen}}},
  \bibinfo{author}{\bibfnamefont{R.-Q.} \bibnamefont{{Xing}}},
  \bibinfo{author}{\bibfnamefont{M.}~\bibnamefont{{Khodas}}}, \bibnamefont{and}
  \bibinfo{author}{\bibfnamefont{A.~V.} \bibnamefont{{Chubukov}}},
  \bibinfo{journal}{ArXiv e-prints}  (\bibinfo{year}{2016}),
  \eprint{1612.08708}.

\bibitem[{\citenamefont{Gretarsson et~al.}(2011)\citenamefont{Gretarsson,
  Lupascu, Kim, Casa, Gog, Wu, Julian, Xu, Wen, Gu et~al.}}]{Gretarsson11}
\bibinfo{author}{\bibfnamefont{H.}~\bibnamefont{Gretarsson}},
  \bibinfo{author}{\bibfnamefont{A.}~\bibnamefont{Lupascu}},
  \bibinfo{author}{\bibfnamefont{J.}~\bibnamefont{Kim}},
  \bibinfo{author}{\bibfnamefont{D.}~\bibnamefont{Casa}},
  \bibinfo{author}{\bibfnamefont{T.}~\bibnamefont{Gog}},
  \bibinfo{author}{\bibfnamefont{W.}~\bibnamefont{Wu}},
  \bibinfo{author}{\bibfnamefont{S.~R.} \bibnamefont{Julian}},
  \bibinfo{author}{\bibfnamefont{Z.~J.} \bibnamefont{Xu}},
  \bibinfo{author}{\bibfnamefont{J.~S.} \bibnamefont{Wen}},
  \bibinfo{author}{\bibfnamefont{G.~D.} \bibnamefont{Gu}},
  \bibnamefont{et~al.}, \bibinfo{journal}{Phys. Rev. B}
  \textbf{\bibinfo{volume}{84}}, \bibinfo{pages}{100509}
  (\bibinfo{year}{2011}),
  \urlprefix\url{http://link.aps.org/doi/10.1103/PhysRevB.84.100509}.

\bibitem[{\citenamefont{Yu and Si}(2015)}]{Si15}
\bibinfo{author}{\bibfnamefont{R.}~\bibnamefont{Yu}} \bibnamefont{and}
  \bibinfo{author}{\bibfnamefont{Q.}~\bibnamefont{Si}}, \bibinfo{journal}{Phys.
  Rev. Lett.} \textbf{\bibinfo{volume}{115}}, \bibinfo{pages}{116401}
  (\bibinfo{year}{2015}),
  \urlprefix\url{http://link.aps.org/doi/10.1103/PhysRevLett.115.116401}.

\bibitem[{\citenamefont{Wang et~al.}(2016{\natexlab{b}})\citenamefont{Wang, Hu,
  and Nevidomskyy}}]{Nevidomskyy16}
\bibinfo{author}{\bibfnamefont{Z.}~\bibnamefont{Wang}},
  \bibinfo{author}{\bibfnamefont{W.-J.} \bibnamefont{Hu}}, \bibnamefont{and}
  \bibinfo{author}{\bibfnamefont{A.~H.} \bibnamefont{Nevidomskyy}},
  \bibinfo{journal}{Phys. Rev. Lett.} \textbf{\bibinfo{volume}{116}},
  \bibinfo{pages}{247203} (\bibinfo{year}{2016}{\natexlab{b}}),
  \urlprefix\url{http://link.aps.org/doi/10.1103/PhysRevLett.116.247203}.

\bibitem[{\citenamefont{Wang et~al.}(2015)\citenamefont{Wang, Kivelson, and
  Lee}}]{FWang15}
\bibinfo{author}{\bibfnamefont{F.}~\bibnamefont{Wang}},
  \bibinfo{author}{\bibfnamefont{S.~A.} \bibnamefont{Kivelson}},
  \bibnamefont{and} \bibinfo{author}{\bibfnamefont{D.-H.} \bibnamefont{Lee}},
  \bibinfo{journal}{Nat Phys} \textbf{\bibinfo{volume}{11}},
  \bibinfo{pages}{959} (\bibinfo{year}{2015}),
  \urlprefix\url{http://dx.doi.org/10.1038/nphys3456}.

\bibitem[{\citenamefont{Affleck et~al.}(1987)\citenamefont{Affleck, Kennedy,
  Lieb, and Tasaki}}]{AKLT}
\bibinfo{author}{\bibfnamefont{I.}~\bibnamefont{Affleck}},
  \bibinfo{author}{\bibfnamefont{T.}~\bibnamefont{Kennedy}},
  \bibinfo{author}{\bibfnamefont{E.~H.} \bibnamefont{Lieb}}, \bibnamefont{and}
  \bibinfo{author}{\bibfnamefont{H.}~\bibnamefont{Tasaki}},
  \bibinfo{journal}{Phys. Rev. Lett.} \textbf{\bibinfo{volume}{59}},
  \bibinfo{pages}{799} (\bibinfo{year}{1987}),
  \urlprefix\url{https://link.aps.org/doi/10.1103/PhysRevLett.59.799}.

\bibitem[{\citenamefont{Glasbrenner et~al.}(2015)\citenamefont{Glasbrenner,
  Mazin, Jeschke, Hirschfeld, Fernandes, and Valenti}}]{Mazin15}
\bibinfo{author}{\bibfnamefont{J.~K.} \bibnamefont{Glasbrenner}},
  \bibinfo{author}{\bibfnamefont{I.~I.} \bibnamefont{Mazin}},
  \bibinfo{author}{\bibfnamefont{H.~O.} \bibnamefont{Jeschke}},
  \bibinfo{author}{\bibfnamefont{P.~J.} \bibnamefont{Hirschfeld}},
  \bibinfo{author}{\bibfnamefont{R.~M.} \bibnamefont{Fernandes}},
  \bibnamefont{and} \bibinfo{author}{\bibfnamefont{R.}~\bibnamefont{Valenti}},
  \bibinfo{journal}{Nat Phys} \textbf{\bibinfo{volume}{11}},
  \bibinfo{pages}{953} (\bibinfo{year}{2015}),
  \urlprefix\url{http://dx.doi.org/10.1038/nphys3434}.

\bibitem[{\citenamefont{Chandra et~al.}(1990)\citenamefont{Chandra, Coleman,
  and Larkin}}]{Chandra90}
\bibinfo{author}{\bibfnamefont{P.}~\bibnamefont{Chandra}},
  \bibinfo{author}{\bibfnamefont{P.}~\bibnamefont{Coleman}}, \bibnamefont{and}
  \bibinfo{author}{\bibfnamefont{A.~I.} \bibnamefont{Larkin}},
  \bibinfo{journal}{Phys. Rev. Lett.} \textbf{\bibinfo{volume}{64}},
  \bibinfo{pages}{88} (\bibinfo{year}{1990}),
  \urlprefix\url{http://link.aps.org/doi/10.1103/PhysRevLett.64.88}.

\bibitem[{\citenamefont{{Misguich} and {Lhuillier}}(2004)}]{Misguich04}
\bibinfo{author}{\bibfnamefont{G.}~\bibnamefont{{Misguich}}} \bibnamefont{and}
  \bibinfo{author}{\bibfnamefont{C.}~\bibnamefont{{Lhuillier}}},
  \emph{\bibinfo{title}{{Two-Dimensional Quantum Antiferromagnets}}}
  (\bibinfo{publisher}{World Scientific Publishing Co}, \bibinfo{year}{2004}),
  pp. \bibinfo{pages}{229--306}.

\bibitem[{\citenamefont{Fernandes et~al.}(2010)\citenamefont{Fernandes,
  VanBebber, Bhattacharya, Chandra, Keppens, Mandrus, McGuire, Sales, Sefat,
  and Schmalian}}]{Fernandes10}
\bibinfo{author}{\bibfnamefont{R.~M.} \bibnamefont{Fernandes}},
  \bibinfo{author}{\bibfnamefont{L.~H.} \bibnamefont{VanBebber}},
  \bibinfo{author}{\bibfnamefont{S.}~\bibnamefont{Bhattacharya}},
  \bibinfo{author}{\bibfnamefont{P.}~\bibnamefont{Chandra}},
  \bibinfo{author}{\bibfnamefont{V.}~\bibnamefont{Keppens}},
  \bibinfo{author}{\bibfnamefont{D.}~\bibnamefont{Mandrus}},
  \bibinfo{author}{\bibfnamefont{M.~A.} \bibnamefont{McGuire}},
  \bibinfo{author}{\bibfnamefont{B.~C.} \bibnamefont{Sales}},
  \bibinfo{author}{\bibfnamefont{A.~S.} \bibnamefont{Sefat}}, \bibnamefont{and}
  \bibinfo{author}{\bibfnamefont{J.}~\bibnamefont{Schmalian}},
  \bibinfo{journal}{Phys. Rev. Lett.} \textbf{\bibinfo{volume}{105}},
  \bibinfo{pages}{157003} (\bibinfo{year}{2010}),
  \urlprefix\url{https://link.aps.org/doi/10.1103/PhysRevLett.105.157003}.

\bibitem[{\citenamefont{Haldane}(1988)}]{Haldane88}
\bibinfo{author}{\bibfnamefont{F.~D.~M.} \bibnamefont{Haldane}},
  \bibinfo{journal}{Phys. Rev. Lett.} \textbf{\bibinfo{volume}{61}},
  \bibinfo{pages}{1029} (\bibinfo{year}{1988}),
  \urlprefix\url{http://link.aps.org/doi/10.1103/PhysRevLett.61.1029}.

\bibitem[{\citenamefont{Read and Sachdev}(1990)}]{Read90}
\bibinfo{author}{\bibfnamefont{N.}~\bibnamefont{Read}} \bibnamefont{and}
  \bibinfo{author}{\bibfnamefont{S.}~\bibnamefont{Sachdev}},
  \bibinfo{journal}{Phys. Rev. B} \textbf{\bibinfo{volume}{42}},
  \bibinfo{pages}{4568} (\bibinfo{year}{1990}),
  \urlprefix\url{http://link.aps.org/doi/10.1103/PhysRevB.42.4568}.

\bibitem[{\citenamefont{Jiang et~al.}(2009)\citenamefont{Jiang, Kr\"uger,
  Moore, Sheng, Zaanen, and Weng}}]{Jiang09}
\bibinfo{author}{\bibfnamefont{H.~C.} \bibnamefont{Jiang}},
  \bibinfo{author}{\bibfnamefont{F.}~\bibnamefont{Kr\"uger}},
  \bibinfo{author}{\bibfnamefont{J.~E.} \bibnamefont{Moore}},
  \bibinfo{author}{\bibfnamefont{D.~N.} \bibnamefont{Sheng}},
  \bibinfo{author}{\bibfnamefont{J.}~\bibnamefont{Zaanen}}, \bibnamefont{and}
  \bibinfo{author}{\bibfnamefont{Z.~Y.} \bibnamefont{Weng}},
  \bibinfo{journal}{Phys. Rev. B} \textbf{\bibinfo{volume}{79}},
  \bibinfo{pages}{174409} (\bibinfo{year}{2009}),
  \urlprefix\url{http://link.aps.org/doi/10.1103/PhysRevB.79.174409}.

\bibitem[{\citenamefont{Jiang et~al.}(2012)\citenamefont{Jiang, Yao, and
  Balents}}]{Jiang12}
\bibinfo{author}{\bibfnamefont{H.-C.} \bibnamefont{Jiang}},
  \bibinfo{author}{\bibfnamefont{H.}~\bibnamefont{Yao}}, \bibnamefont{and}
  \bibinfo{author}{\bibfnamefont{L.}~\bibnamefont{Balents}},
  \bibinfo{journal}{Phys. Rev. B} \textbf{\bibinfo{volume}{86}},
  \bibinfo{pages}{024424} (\bibinfo{year}{2012}),
  \urlprefix\url{http://link.aps.org/doi/10.1103/PhysRevB.86.024424}.

\bibitem[{\citenamefont{{Gong} et~al.}(2016)\citenamefont{{Gong}, {Zhu},
  {Sheng}, and {Yang}}}]{KYang16}
\bibinfo{author}{\bibfnamefont{S.-S.} \bibnamefont{{Gong}}},
  \bibinfo{author}{\bibfnamefont{W.}~\bibnamefont{{Zhu}}},
  \bibinfo{author}{\bibfnamefont{D.~N.} \bibnamefont{{Sheng}}},
  \bibnamefont{and} \bibinfo{author}{\bibfnamefont{K.}~\bibnamefont{{Yang}}},
  \bibinfo{journal}{ArXiv e-prints}  (\bibinfo{year}{2016}),
  \eprint{1606.00937}.

\bibitem[{\citenamefont{Nakayama et~al.}(2014)\citenamefont{Nakayama, Miyata,
  Phan, Sato, Tanabe, Urata, Tanigaki, and Takahashi}}]{Nakayama14}
\bibinfo{author}{\bibfnamefont{K.}~\bibnamefont{Nakayama}},
  \bibinfo{author}{\bibfnamefont{Y.}~\bibnamefont{Miyata}},
  \bibinfo{author}{\bibfnamefont{G.~N.} \bibnamefont{Phan}},
  \bibinfo{author}{\bibfnamefont{T.}~\bibnamefont{Sato}},
  \bibinfo{author}{\bibfnamefont{Y.}~\bibnamefont{Tanabe}},
  \bibinfo{author}{\bibfnamefont{T.}~\bibnamefont{Urata}},
  \bibinfo{author}{\bibfnamefont{K.}~\bibnamefont{Tanigaki}}, \bibnamefont{and}
  \bibinfo{author}{\bibfnamefont{T.}~\bibnamefont{Takahashi}},
  \bibinfo{journal}{Phys. Rev. Lett.} \textbf{\bibinfo{volume}{113}},
  \bibinfo{pages}{237001} (\bibinfo{year}{2014}),
  \urlprefix\url{http://link.aps.org/doi/10.1103/PhysRevLett.113.237001}.

\bibitem[{\citenamefont{Watson et~al.}(2015)\citenamefont{Watson, Kim,
  Haghighirad, Davies, McCollam, Narayanan, Blake, Chen, Ghannadzadeh,
  Schofield et~al.}}]{Watson15}
\bibinfo{author}{\bibfnamefont{M.~D.} \bibnamefont{Watson}},
  \bibinfo{author}{\bibfnamefont{T.~K.} \bibnamefont{Kim}},
  \bibinfo{author}{\bibfnamefont{A.~A.} \bibnamefont{Haghighirad}},
  \bibinfo{author}{\bibfnamefont{N.~R.} \bibnamefont{Davies}},
  \bibinfo{author}{\bibfnamefont{A.}~\bibnamefont{McCollam}},
  \bibinfo{author}{\bibfnamefont{A.}~\bibnamefont{Narayanan}},
  \bibinfo{author}{\bibfnamefont{S.~F.} \bibnamefont{Blake}},
  \bibinfo{author}{\bibfnamefont{Y.~L.} \bibnamefont{Chen}},
  \bibinfo{author}{\bibfnamefont{S.}~\bibnamefont{Ghannadzadeh}},
  \bibinfo{author}{\bibfnamefont{A.~J.} \bibnamefont{Schofield}},
  \bibnamefont{et~al.}, \bibinfo{journal}{Phys. Rev. B}
  \textbf{\bibinfo{volume}{91}}, \bibinfo{pages}{155106}
  (\bibinfo{year}{2015}),
  \urlprefix\url{http://link.aps.org/doi/10.1103/PhysRevB.91.155106}.

\bibitem[{\citenamefont{Suzuki et~al.}(2015)\citenamefont{Suzuki, Shimojima,
  Sonobe, Nakamura, Sakano, Tsuji, Omachi, Yoshioka, Kuwata-Gonokami, Watashige
  et~al.}}]{Suzuki15}
\bibinfo{author}{\bibfnamefont{Y.}~\bibnamefont{Suzuki}},
  \bibinfo{author}{\bibfnamefont{T.}~\bibnamefont{Shimojima}},
  \bibinfo{author}{\bibfnamefont{T.}~\bibnamefont{Sonobe}},
  \bibinfo{author}{\bibfnamefont{A.}~\bibnamefont{Nakamura}},
  \bibinfo{author}{\bibfnamefont{M.}~\bibnamefont{Sakano}},
  \bibinfo{author}{\bibfnamefont{H.}~\bibnamefont{Tsuji}},
  \bibinfo{author}{\bibfnamefont{J.}~\bibnamefont{Omachi}},
  \bibinfo{author}{\bibfnamefont{K.}~\bibnamefont{Yoshioka}},
  \bibinfo{author}{\bibfnamefont{M.}~\bibnamefont{Kuwata-Gonokami}},
  \bibinfo{author}{\bibfnamefont{T.}~\bibnamefont{Watashige}},
  \bibnamefont{et~al.}, \bibinfo{journal}{Phys. Rev. B}
  \textbf{\bibinfo{volume}{92}}, \bibinfo{pages}{205117}
  (\bibinfo{year}{2015}),
  \urlprefix\url{http://link.aps.org/doi/10.1103/PhysRevB.92.205117}.

\bibitem[{\citenamefont{Moon et~al.}(2010)\citenamefont{Moon, Shin, Parker,
  Choi, Mazin, Lee, Kim, Sung, Cho, Khim et~al.}}]{Moon10}
\bibinfo{author}{\bibfnamefont{S.~J.} \bibnamefont{Moon}},
  \bibinfo{author}{\bibfnamefont{J.~H.} \bibnamefont{Shin}},
  \bibinfo{author}{\bibfnamefont{D.}~\bibnamefont{Parker}},
  \bibinfo{author}{\bibfnamefont{W.~S.} \bibnamefont{Choi}},
  \bibinfo{author}{\bibfnamefont{I.~I.} \bibnamefont{Mazin}},
  \bibinfo{author}{\bibfnamefont{Y.~S.} \bibnamefont{Lee}},
  \bibinfo{author}{\bibfnamefont{J.~Y.} \bibnamefont{Kim}},
  \bibinfo{author}{\bibfnamefont{N.~H.} \bibnamefont{Sung}},
  \bibinfo{author}{\bibfnamefont{B.~K.} \bibnamefont{Cho}},
  \bibinfo{author}{\bibfnamefont{S.~H.} \bibnamefont{Khim}},
  \bibnamefont{et~al.}, \bibinfo{journal}{Phys. Rev. B}
  \textbf{\bibinfo{volume}{81}}, \bibinfo{pages}{205114}
  (\bibinfo{year}{2010}),
  \urlprefix\url{http://link.aps.org/doi/10.1103/PhysRevB.81.205114}.

\bibitem[{\citenamefont{Kou et~al.}(2009)\citenamefont{Kou, Li, and
  Weng}}]{Weng09}
\bibinfo{author}{\bibfnamefont{S.-P.} \bibnamefont{Kou}},
  \bibinfo{author}{\bibfnamefont{T.}~\bibnamefont{Li}}, \bibnamefont{and}
  \bibinfo{author}{\bibfnamefont{Z.-Y.} \bibnamefont{Weng}},
  \bibinfo{journal}{EPL (Europhysics Letters)} \textbf{\bibinfo{volume}{88}},
  \bibinfo{pages}{17010} (\bibinfo{year}{2009}),
  \urlprefix\url{http://stacks.iop.org/0295-5075/88/i=1/a=17010}.

\bibitem[{\citenamefont{Lv et~al.}(2010)\citenamefont{Lv, Kr\"uger, and
  Phillips}}]{Phillips10}
\bibinfo{author}{\bibfnamefont{W.}~\bibnamefont{Lv}},
  \bibinfo{author}{\bibfnamefont{F.}~\bibnamefont{Kr\"uger}}, \bibnamefont{and}
  \bibinfo{author}{\bibfnamefont{P.}~\bibnamefont{Phillips}},
  \bibinfo{journal}{Phys. Rev. B} \textbf{\bibinfo{volume}{82}},
  \bibinfo{pages}{045125} (\bibinfo{year}{2010}),
  \urlprefix\url{http://link.aps.org/doi/10.1103/PhysRevB.82.045125}.

\bibitem[{\citenamefont{Yin et~al.}(2010)\citenamefont{Yin, Lee, and
  Ku}}]{Ku10}
\bibinfo{author}{\bibfnamefont{W.-G.} \bibnamefont{Yin}},
  \bibinfo{author}{\bibfnamefont{C.-C.} \bibnamefont{Lee}}, \bibnamefont{and}
  \bibinfo{author}{\bibfnamefont{W.}~\bibnamefont{Ku}}, \bibinfo{journal}{Phys.
  Rev. Lett.} \textbf{\bibinfo{volume}{105}}, \bibinfo{pages}{107004}
  (\bibinfo{year}{2010}),
  \urlprefix\url{http://link.aps.org/doi/10.1103/PhysRevLett.105.107004}.

\bibitem[{\citenamefont{Stanev and Littlewood}(2013)}]{Littlewood13}
\bibinfo{author}{\bibfnamefont{V.}~\bibnamefont{Stanev}} \bibnamefont{and}
  \bibinfo{author}{\bibfnamefont{P.~B.} \bibnamefont{Littlewood}},
  \bibinfo{journal}{Phys. Rev. B} \textbf{\bibinfo{volume}{87}},
  \bibinfo{pages}{161122} (\bibinfo{year}{2013}),
  \urlprefix\url{http://link.aps.org/doi/10.1103/PhysRevB.87.161122}.

\bibitem[{\citenamefont{Liang et~al.}(2014)\citenamefont{Liang, Mukherjee,
  Patel, Bishop, Dagotto, and Moreo}}]{Dagotto14}
\bibinfo{author}{\bibfnamefont{S.}~\bibnamefont{Liang}},
  \bibinfo{author}{\bibfnamefont{A.}~\bibnamefont{Mukherjee}},
  \bibinfo{author}{\bibfnamefont{N.~D.} \bibnamefont{Patel}},
  \bibinfo{author}{\bibfnamefont{C.~B.} \bibnamefont{Bishop}},
  \bibinfo{author}{\bibfnamefont{E.}~\bibnamefont{Dagotto}}, \bibnamefont{and}
  \bibinfo{author}{\bibfnamefont{A.}~\bibnamefont{Moreo}},
  \bibinfo{journal}{Phys. Rev. B} \textbf{\bibinfo{volume}{90}},
  \bibinfo{pages}{184507} (\bibinfo{year}{2014}),
  \urlprefix\url{http://link.aps.org/doi/10.1103/PhysRevB.90.184507}.

\bibitem[{\citenamefont{Lee}(2015)}]{DHLee15}
\bibinfo{author}{\bibfnamefont{D.-H.} \bibnamefont{Lee}},
  \bibinfo{journal}{Chinese Physics B} \textbf{\bibinfo{volume}{24}},
  \bibinfo{pages}{117405} (\bibinfo{year}{2015}),
  \urlprefix\url{http://stacks.iop.org/1674-1056/24/i=11/a=117405}.

\bibitem[{\citenamefont{Fang et~al.}(2008)\citenamefont{Fang, Yao, Tsai, Hu,
  and Kivelson}}]{Chen08}
\bibinfo{author}{\bibfnamefont{C.}~\bibnamefont{Fang}},
  \bibinfo{author}{\bibfnamefont{H.}~\bibnamefont{Yao}},
  \bibinfo{author}{\bibfnamefont{W.-F.} \bibnamefont{Tsai}},
  \bibinfo{author}{\bibfnamefont{J.}~\bibnamefont{Hu}}, \bibnamefont{and}
  \bibinfo{author}{\bibfnamefont{S.~A.} \bibnamefont{Kivelson}},
  \bibinfo{journal}{Phys. Rev. B} \textbf{\bibinfo{volume}{77}},
  \bibinfo{pages}{224509} (\bibinfo{year}{2008}),
  \urlprefix\url{http://link.aps.org/doi/10.1103/PhysRevB.77.224509}.

\bibitem[{\citenamefont{Xu et~al.}(2008)\citenamefont{Xu, M\"uller, and
  Sachdev}}]{Xu08}
\bibinfo{author}{\bibfnamefont{C.}~\bibnamefont{Xu}},
  \bibinfo{author}{\bibfnamefont{M.}~\bibnamefont{M\"uller}}, \bibnamefont{and}
  \bibinfo{author}{\bibfnamefont{S.}~\bibnamefont{Sachdev}},
  \bibinfo{journal}{Phys. Rev. B} \textbf{\bibinfo{volume}{78}},
  \bibinfo{pages}{020501} (\bibinfo{year}{2008}),
  \urlprefix\url{http://link.aps.org/doi/10.1103/PhysRevB.78.020501}.

\bibitem[{\citenamefont{Balents}(2010)}]{Balents10}
\bibinfo{author}{\bibfnamefont{L.}~\bibnamefont{Balents}},
  \bibinfo{journal}{Nature} \textbf{\bibinfo{volume}{464}},
  \bibinfo{pages}{199} (\bibinfo{year}{2010}).

\bibitem[{\citenamefont{Arovas and Auerbach}(1988)}]{Arovas88}
\bibinfo{author}{\bibfnamefont{D.~P.} \bibnamefont{Arovas}} \bibnamefont{and}
  \bibinfo{author}{\bibfnamefont{A.}~\bibnamefont{Auerbach}},
  \bibinfo{journal}{Phys. Rev. B} \textbf{\bibinfo{volume}{38}},
  \bibinfo{pages}{316} (\bibinfo{year}{1988}),
  \urlprefix\url{http://link.aps.org/doi/10.1103/PhysRevB.38.316}.

\bibitem[{\citenamefont{M\"uller et~al.}(1981)\citenamefont{M\"uller, Thomas,
  Beck, and Bonner}}]{Muller81}
\bibinfo{author}{\bibfnamefont{G.}~\bibnamefont{M\"uller}},
  \bibinfo{author}{\bibfnamefont{H.}~\bibnamefont{Thomas}},
  \bibinfo{author}{\bibfnamefont{H.}~\bibnamefont{Beck}}, \bibnamefont{and}
  \bibinfo{author}{\bibfnamefont{J.~C.} \bibnamefont{Bonner}},
  \bibinfo{journal}{Phys. Rev. B} \textbf{\bibinfo{volume}{24}},
  \bibinfo{pages}{1429} (\bibinfo{year}{1981}),
  \urlprefix\url{http://link.aps.org/doi/10.1103/PhysRevB.24.1429}.

\bibitem[{\citenamefont{Lake et~al.}(2013)\citenamefont{Lake, Tennant, Caux,
  Barthel, Schollw\"ock, Nagler, and Frost}}]{Lake13}
\bibinfo{author}{\bibfnamefont{B.}~\bibnamefont{Lake}},
  \bibinfo{author}{\bibfnamefont{D.~A.} \bibnamefont{Tennant}},
  \bibinfo{author}{\bibfnamefont{J.-S.} \bibnamefont{Caux}},
  \bibinfo{author}{\bibfnamefont{T.}~\bibnamefont{Barthel}},
  \bibinfo{author}{\bibfnamefont{U.}~\bibnamefont{Schollw\"ock}},
  \bibinfo{author}{\bibfnamefont{S.~E.} \bibnamefont{Nagler}},
  \bibnamefont{and} \bibinfo{author}{\bibfnamefont{C.~D.} \bibnamefont{Frost}},
  \bibinfo{journal}{Phys. Rev. Lett.} \textbf{\bibinfo{volume}{111}},
  \bibinfo{pages}{137205} (\bibinfo{year}{2013}),
  \urlprefix\url{http://link.aps.org/doi/10.1103/PhysRevLett.111.137205}.

\bibitem[{\citenamefont{{Vlijm} and {Caux}}(2014)}]{Caux14}
\bibinfo{author}{\bibfnamefont{R.}~\bibnamefont{{Vlijm}}} \bibnamefont{and}
  \bibinfo{author}{\bibfnamefont{J.-S.} \bibnamefont{{Caux}}},
  \bibinfo{journal}{Journal of Statistical Mechanics: Theory and Experiment}
  \textbf{\bibinfo{volume}{5}}, \bibinfo{eid}{05009} (\bibinfo{year}{2014}),
  \eprint{1401.4450}.

\bibitem[{\citenamefont{{Sachdev} and {Read}}(1991)}]{Sachdev91}
\bibinfo{author}{\bibfnamefont{S.}~\bibnamefont{{Sachdev}}} \bibnamefont{and}
  \bibinfo{author}{\bibfnamefont{N.}~\bibnamefont{{Read}}},
  \bibinfo{journal}{International Journal of Modern Physics B}
  \textbf{\bibinfo{volume}{5}}, \bibinfo{pages}{219} (\bibinfo{year}{1991}),
  \eprint{cond-mat/0402109}.

\bibitem[{\citenamefont{{Tchernyshyov}
  et~al.}(2006)\citenamefont{{Tchernyshyov}, {Moessner}, and
  {Sondhi}}}]{Tchernyshyov06}
\bibinfo{author}{\bibfnamefont{O.}~\bibnamefont{{Tchernyshyov}}},
  \bibinfo{author}{\bibfnamefont{R.}~\bibnamefont{{Moessner}}},
  \bibnamefont{and} \bibinfo{author}{\bibfnamefont{S.~L.}
  \bibnamefont{{Sondhi}}}, \bibinfo{journal}{EPL (Europhysics Letters)}
  \textbf{\bibinfo{volume}{73}}, \bibinfo{pages}{278} (\bibinfo{year}{2006}),
  \eprint{cond-mat/0408498}.

\bibitem[{\citenamefont{Auerbach and Arovas}(1988)}]{Auerbach88}
\bibinfo{author}{\bibfnamefont{A.}~\bibnamefont{Auerbach}} \bibnamefont{and}
  \bibinfo{author}{\bibfnamefont{D.~P.} \bibnamefont{Arovas}},
  \bibinfo{journal}{Phys. Rev. Lett.} \textbf{\bibinfo{volume}{61}},
  \bibinfo{pages}{617} (\bibinfo{year}{1988}),
  \urlprefix\url{http://link.aps.org/doi/10.1103/PhysRevLett.61.617}.

\bibitem[{\citenamefont{Cvetkovic and Vafek}(2013)}]{Vafek13}
\bibinfo{author}{\bibfnamefont{V.}~\bibnamefont{Cvetkovic}} \bibnamefont{and}
  \bibinfo{author}{\bibfnamefont{O.}~\bibnamefont{Vafek}},
  \bibinfo{journal}{Phys. Rev. B} \textbf{\bibinfo{volume}{88}},
  \bibinfo{pages}{134510} (\bibinfo{year}{2013}),
  \urlprefix\url{http://link.aps.org/doi/10.1103/PhysRevB.88.134510}.

\bibitem[{\citenamefont{Baek et~al.}(2015)\citenamefont{Baek, Efremov, Ok, Kim,
  van~den Brink, and B{\"u}chner}}]{Brink15}
\bibinfo{author}{\bibfnamefont{S.-H.} \bibnamefont{Baek}},
  \bibinfo{author}{\bibfnamefont{D.~V.} \bibnamefont{Efremov}},
  \bibinfo{author}{\bibfnamefont{J.~M.} \bibnamefont{Ok}},
  \bibinfo{author}{\bibfnamefont{J.~S.} \bibnamefont{Kim}},
  \bibinfo{author}{\bibfnamefont{J.}~\bibnamefont{van~den Brink}},
  \bibnamefont{and}
  \bibinfo{author}{\bibfnamefont{B.}~\bibnamefont{B{\"u}chner}},
  \bibinfo{journal}{Nat Mater} \textbf{\bibinfo{volume}{14}},
  \bibinfo{pages}{210} (\bibinfo{year}{2015}),
  \urlprefix\url{http://dx.doi.org/10.1038/nmat4138}.

\bibitem[{\citenamefont{{Sprau} et~al.}(2016)\citenamefont{{Sprau}, {Kostin},
  {Kreisel}, {B{\"o}hmer}, {Taufour}, {Canfield}, {Mukherjee}, {Hirschfeld},
  {Andersen}, and {S{\'e}amus Davis}}}]{Sprau16}
\bibinfo{author}{\bibfnamefont{P.~O.} \bibnamefont{{Sprau}}},
  \bibinfo{author}{\bibfnamefont{A.}~\bibnamefont{{Kostin}}},
  \bibinfo{author}{\bibfnamefont{A.}~\bibnamefont{{Kreisel}}},
  \bibinfo{author}{\bibfnamefont{A.~E.} \bibnamefont{{B{\"o}hmer}}},
  \bibinfo{author}{\bibfnamefont{V.}~\bibnamefont{{Taufour}}},
  \bibinfo{author}{\bibfnamefont{P.~C.} \bibnamefont{{Canfield}}},
  \bibinfo{author}{\bibfnamefont{S.}~\bibnamefont{{Mukherjee}}},
  \bibinfo{author}{\bibfnamefont{P.~J.} \bibnamefont{{Hirschfeld}}},
  \bibinfo{author}{\bibfnamefont{B.~M.} \bibnamefont{{Andersen}}},
  \bibnamefont{and} \bibinfo{author}{\bibfnamefont{J.~C.}
  \bibnamefont{{S{\'e}amus Davis}}}, \bibinfo{journal}{ArXiv e-prints}
  (\bibinfo{year}{2016}), \eprint{1611.02134}.

\bibitem[{\citenamefont{Song et~al.}(2011)\citenamefont{Song, Wang, Cheng,
  Jiang, Li, Zhang, Li, He, Wang, Jia et~al.}}]{QKXue11}
\bibinfo{author}{\bibfnamefont{C.-L.} \bibnamefont{Song}},
  \bibinfo{author}{\bibfnamefont{Y.-L.} \bibnamefont{Wang}},
  \bibinfo{author}{\bibfnamefont{P.}~\bibnamefont{Cheng}},
  \bibinfo{author}{\bibfnamefont{Y.-P.} \bibnamefont{Jiang}},
  \bibinfo{author}{\bibfnamefont{W.}~\bibnamefont{Li}},
  \bibinfo{author}{\bibfnamefont{T.}~\bibnamefont{Zhang}},
  \bibinfo{author}{\bibfnamefont{Z.}~\bibnamefont{Li}},
  \bibinfo{author}{\bibfnamefont{K.}~\bibnamefont{He}},
  \bibinfo{author}{\bibfnamefont{L.}~\bibnamefont{Wang}},
  \bibinfo{author}{\bibfnamefont{J.-F.} \bibnamefont{Jia}},
  \bibnamefont{et~al.}, \bibinfo{journal}{Science}
  \textbf{\bibinfo{volume}{332}}, \bibinfo{pages}{1410} (\bibinfo{year}{2011}),
  ISSN \bibinfo{issn}{0036-8075},
  \eprint{http://science.sciencemag.org/content/332/6036/1410.full.pdf},
  \urlprefix\url{http://science.sciencemag.org/content/332/6036/1410}.

\bibitem[{\citenamefont{Kasahara
  et~al.}(2014{\natexlab{a}})\citenamefont{Kasahara, Watashige, Hanaguri,
  Kohsaka, Yamashita, Shimoyama, Mizukami, Endo, Ikeda, Aoyama
  et~al.}}]{Matsuda14}
\bibinfo{author}{\bibfnamefont{S.}~\bibnamefont{Kasahara}},
  \bibinfo{author}{\bibfnamefont{T.}~\bibnamefont{Watashige}},
  \bibinfo{author}{\bibfnamefont{T.}~\bibnamefont{Hanaguri}},
  \bibinfo{author}{\bibfnamefont{Y.}~\bibnamefont{Kohsaka}},
  \bibinfo{author}{\bibfnamefont{T.}~\bibnamefont{Yamashita}},
  \bibinfo{author}{\bibfnamefont{Y.}~\bibnamefont{Shimoyama}},
  \bibinfo{author}{\bibfnamefont{Y.}~\bibnamefont{Mizukami}},
  \bibinfo{author}{\bibfnamefont{R.}~\bibnamefont{Endo}},
  \bibinfo{author}{\bibfnamefont{H.}~\bibnamefont{Ikeda}},
  \bibinfo{author}{\bibfnamefont{K.}~\bibnamefont{Aoyama}},
  \bibnamefont{et~al.}, \bibinfo{journal}{Proceedings of the National Academy
  of Sciences} \textbf{\bibinfo{volume}{111}}, \bibinfo{pages}{16309}
  (\bibinfo{year}{2014}{\natexlab{a}}),
  \eprint{http://www.pnas.org/content/111/46/16309.full.pdf},
  \urlprefix\url{http://www.pnas.org/content/111/46/16309.abstract}.

\bibitem[{\citenamefont{Bourgeois-Hope
  et~al.}(2016)\citenamefont{Bourgeois-Hope, Chi, Bonn, Liang, Hardy, Wolf,
  Meingast, Doiron-Leyraud, and Taillefer}}]{Taillefer16}
\bibinfo{author}{\bibfnamefont{P.}~\bibnamefont{Bourgeois-Hope}},
  \bibinfo{author}{\bibfnamefont{S.}~\bibnamefont{Chi}},
  \bibinfo{author}{\bibfnamefont{D.~A.} \bibnamefont{Bonn}},
  \bibinfo{author}{\bibfnamefont{R.}~\bibnamefont{Liang}},
  \bibinfo{author}{\bibfnamefont{W.~N.} \bibnamefont{Hardy}},
  \bibinfo{author}{\bibfnamefont{T.}~\bibnamefont{Wolf}},
  \bibinfo{author}{\bibfnamefont{C.}~\bibnamefont{Meingast}},
  \bibinfo{author}{\bibfnamefont{N.}~\bibnamefont{Doiron-Leyraud}},
  \bibnamefont{and}
  \bibinfo{author}{\bibfnamefont{L.}~\bibnamefont{Taillefer}},
  \bibinfo{journal}{Phys. Rev. Lett.} \textbf{\bibinfo{volume}{117}},
  \bibinfo{pages}{097003} (\bibinfo{year}{2016}),
  \urlprefix\url{http://link.aps.org/doi/10.1103/PhysRevLett.117.097003}.

\bibitem[{\citenamefont{{Wang} et~al.}(2016)\citenamefont{{Wang}, {Shen},
  {Pan}, {Hao}, {Ma}, {Zhou}, {Steffens}, {Schmalzl}, {Forrest}, {Abdel-Hafiez}
  et~al.}}]{JZhao15}
\bibinfo{author}{\bibfnamefont{Q.}~\bibnamefont{{Wang}}},
  \bibinfo{author}{\bibfnamefont{Y.}~\bibnamefont{{Shen}}},
  \bibinfo{author}{\bibfnamefont{B.}~\bibnamefont{{Pan}}},
  \bibinfo{author}{\bibfnamefont{Y.}~\bibnamefont{{Hao}}},
  \bibinfo{author}{\bibfnamefont{M.}~\bibnamefont{{Ma}}},
  \bibinfo{author}{\bibfnamefont{F.}~\bibnamefont{{Zhou}}},
  \bibinfo{author}{\bibfnamefont{P.}~\bibnamefont{{Steffens}}},
  \bibinfo{author}{\bibfnamefont{K.}~\bibnamefont{{Schmalzl}}},
  \bibinfo{author}{\bibfnamefont{T.~R.} \bibnamefont{{Forrest}}},
  \bibinfo{author}{\bibfnamefont{M.}~\bibnamefont{{Abdel-Hafiez}}},
  \bibnamefont{et~al.}, \bibinfo{journal}{Nature Materials}
  \textbf{\bibinfo{volume}{15}}, \bibinfo{pages}{159} (\bibinfo{year}{2016}),
  \eprint{1502.07544}.

\bibitem[{\citenamefont{{Yi} et~al.}(2015)\citenamefont{{Yi}, {Liu}, {Zhang},
  {Yu}, {Zhu}, {Lee}, {Moore}, {Schmitt}, {Li}, {Riggs} et~al.}}]{DHLu15}
\bibinfo{author}{\bibfnamefont{M.}~\bibnamefont{{Yi}}},
  \bibinfo{author}{\bibfnamefont{Z.-K.} \bibnamefont{{Liu}}},
  \bibinfo{author}{\bibfnamefont{Y.}~\bibnamefont{{Zhang}}},
  \bibinfo{author}{\bibfnamefont{R.}~\bibnamefont{{Yu}}},
  \bibinfo{author}{\bibfnamefont{J.-X.} \bibnamefont{{Zhu}}},
  \bibinfo{author}{\bibfnamefont{J.~J.} \bibnamefont{{Lee}}},
  \bibinfo{author}{\bibfnamefont{R.~G.} \bibnamefont{{Moore}}},
  \bibinfo{author}{\bibfnamefont{F.~T.} \bibnamefont{{Schmitt}}},
  \bibinfo{author}{\bibfnamefont{W.}~\bibnamefont{{Li}}},
  \bibinfo{author}{\bibfnamefont{S.~C.} \bibnamefont{{Riggs}}},
  \bibnamefont{et~al.}, \bibinfo{journal}{Nature Communications}
  \textbf{\bibinfo{volume}{6}}, \bibinfo{eid}{7777} (\bibinfo{year}{2015}),
  \eprint{1506.03888}.

\bibitem[{\citenamefont{Senthil et~al.}(2003)\citenamefont{Senthil, Sachdev,
  and Vojta}}]{Senthil03}
\bibinfo{author}{\bibfnamefont{T.}~\bibnamefont{Senthil}},
  \bibinfo{author}{\bibfnamefont{S.}~\bibnamefont{Sachdev}}, \bibnamefont{and}
  \bibinfo{author}{\bibfnamefont{M.}~\bibnamefont{Vojta}},
  \bibinfo{journal}{Phys. Rev. Lett.} \textbf{\bibinfo{volume}{90}},
  \bibinfo{pages}{216403} (\bibinfo{year}{2003}),
  \urlprefix\url{http://link.aps.org/doi/10.1103/PhysRevLett.90.216403}.

\bibitem[{\citenamefont{Senthil et~al.}(2004)\citenamefont{Senthil, Vojta, and
  Sachdev}}]{Senthil04}
\bibinfo{author}{\bibfnamefont{T.}~\bibnamefont{Senthil}},
  \bibinfo{author}{\bibfnamefont{M.}~\bibnamefont{Vojta}}, \bibnamefont{and}
  \bibinfo{author}{\bibfnamefont{S.}~\bibnamefont{Sachdev}},
  \bibinfo{journal}{Phys. Rev. B} \textbf{\bibinfo{volume}{69}},
  \bibinfo{pages}{035111} (\bibinfo{year}{2004}),
  \urlprefix\url{http://link.aps.org/doi/10.1103/PhysRevB.69.035111}.

\bibitem[{\citenamefont{Lee et~al.}(2016)\citenamefont{Lee, Sachdev, and
  White}}]{Sachdev16}
\bibinfo{author}{\bibfnamefont{J.}~\bibnamefont{Lee}},
  \bibinfo{author}{\bibfnamefont{S.}~\bibnamefont{Sachdev}}, \bibnamefont{and}
  \bibinfo{author}{\bibfnamefont{S.~R.} \bibnamefont{White}},
  \bibinfo{journal}{Phys. Rev. B} \textbf{\bibinfo{volume}{94}},
  \bibinfo{pages}{115112} (\bibinfo{year}{2016}),
  \urlprefix\url{http://link.aps.org/doi/10.1103/PhysRevB.94.115112}.

\bibitem[{\citenamefont{Kasahara
  et~al.}(2014{\natexlab{b}})\citenamefont{Kasahara, Watashige, Hanaguri,
  Kohsaka, Yamashita, Shimoyama, Mizukami, Endo, Ikeda, Aoyama
  et~al.}}]{Kasahara14}
\bibinfo{author}{\bibfnamefont{S.}~\bibnamefont{Kasahara}},
  \bibinfo{author}{\bibfnamefont{T.}~\bibnamefont{Watashige}},
  \bibinfo{author}{\bibfnamefont{T.}~\bibnamefont{Hanaguri}},
  \bibinfo{author}{\bibfnamefont{Y.}~\bibnamefont{Kohsaka}},
  \bibinfo{author}{\bibfnamefont{T.}~\bibnamefont{Yamashita}},
  \bibinfo{author}{\bibfnamefont{Y.}~\bibnamefont{Shimoyama}},
  \bibinfo{author}{\bibfnamefont{Y.}~\bibnamefont{Mizukami}},
  \bibinfo{author}{\bibfnamefont{R.}~\bibnamefont{Endo}},
  \bibinfo{author}{\bibfnamefont{H.}~\bibnamefont{Ikeda}},
  \bibinfo{author}{\bibfnamefont{K.}~\bibnamefont{Aoyama}},
  \bibnamefont{et~al.}, \bibinfo{journal}{Proceedings of the National Academy
  of Sciences} \textbf{\bibinfo{volume}{111}}, \bibinfo{pages}{16309}
  (\bibinfo{year}{2014}{\natexlab{b}}),
  \eprint{http://www.pnas.org/content/111/46/16309.full.pdf},
  \urlprefix\url{http://www.pnas.org/content/111/46/16309.abstract}.

\bibitem[{\citenamefont{Kirzhnits et~al.}(1973)\citenamefont{Kirzhnits,
  Maksimov, and Khomskii}}]{Kirzhnits1973}
\bibinfo{author}{\bibfnamefont{D.~A.} \bibnamefont{Kirzhnits}},
  \bibinfo{author}{\bibfnamefont{E.~G.} \bibnamefont{Maksimov}},
  \bibnamefont{and} \bibinfo{author}{\bibfnamefont{D.~I.}
  \bibnamefont{Khomskii}}, \bibinfo{journal}{Journal of Low Temperature
  Physics} \textbf{\bibinfo{volume}{10}}, \bibinfo{pages}{79}
  (\bibinfo{year}{1973}), ISSN \bibinfo{issn}{1573-7357},
  \urlprefix\url{http://dx.doi.org/10.1007/BF00655243}.

\bibitem[{\citenamefont{Klimin et~al.}(2014)\citenamefont{Klimin, Tempere,
  Devreese, and van~der Marel}}]{Marel14}
\bibinfo{author}{\bibfnamefont{S.~N.} \bibnamefont{Klimin}},
  \bibinfo{author}{\bibfnamefont{J.}~\bibnamefont{Tempere}},
  \bibinfo{author}{\bibfnamefont{J.~T.} \bibnamefont{Devreese}},
  \bibnamefont{and} \bibinfo{author}{\bibfnamefont{D.}~\bibnamefont{van~der
  Marel}}, \bibinfo{journal}{Phys. Rev. B} \textbf{\bibinfo{volume}{89}},
  \bibinfo{pages}{184514} (\bibinfo{year}{2014}),
  \urlprefix\url{http://link.aps.org/doi/10.1103/PhysRevB.89.184514}.

\bibitem[{\citenamefont{Miyata et~al.}(2015)\citenamefont{Miyata, Nakayama,
  Sugawara, Sato, and Takahashi}}]{Miyata15}
\bibinfo{author}{\bibfnamefont{Y.}~\bibnamefont{Miyata}},
  \bibinfo{author}{\bibfnamefont{K.}~\bibnamefont{Nakayama}},
  \bibinfo{author}{\bibfnamefont{K.}~\bibnamefont{Sugawara}},
  \bibinfo{author}{\bibfnamefont{T.}~\bibnamefont{Sato}}, \bibnamefont{and}
  \bibinfo{author}{\bibfnamefont{T.}~\bibnamefont{Takahashi}},
  \bibinfo{journal}{Nat Mater} \textbf{\bibinfo{volume}{14}},
  \bibinfo{pages}{775} (\bibinfo{year}{2015}),
  \urlprefix\url{http://dx.doi.org/10.1038/nmat4302}.

\bibitem[{\citenamefont{Lee et~al.}(2014)\citenamefont{Lee, Schmitt, Moore,
  Johnston, Cui, Li, Yi, Liu, Hashimoto, Zhang et~al.}}]{ZXShen14}
\bibinfo{author}{\bibfnamefont{J.~J.} \bibnamefont{Lee}},
  \bibinfo{author}{\bibfnamefont{F.~T.} \bibnamefont{Schmitt}},
  \bibinfo{author}{\bibfnamefont{R.~G.} \bibnamefont{Moore}},
  \bibinfo{author}{\bibfnamefont{S.}~\bibnamefont{Johnston}},
  \bibinfo{author}{\bibfnamefont{Y.~T.} \bibnamefont{Cui}},
  \bibinfo{author}{\bibfnamefont{W.}~\bibnamefont{Li}},
  \bibinfo{author}{\bibfnamefont{M.}~\bibnamefont{Yi}},
  \bibinfo{author}{\bibfnamefont{Z.~K.} \bibnamefont{Liu}},
  \bibinfo{author}{\bibfnamefont{M.}~\bibnamefont{Hashimoto}},
  \bibinfo{author}{\bibfnamefont{Y.}~\bibnamefont{Zhang}},
  \bibnamefont{et~al.}, \bibinfo{journal}{Nature}
  \textbf{\bibinfo{volume}{515}}, \bibinfo{pages}{245} (\bibinfo{year}{2014}),
  \urlprefix\url{http://dx.doi.org/10.1038/nature13894}.

\bibitem[{\citenamefont{Bulut et~al.}(1992)\citenamefont{Bulut, Scalapino, and
  Scalettar}}]{Bulut92}
\bibinfo{author}{\bibfnamefont{N.}~\bibnamefont{Bulut}},
  \bibinfo{author}{\bibfnamefont{D.~J.} \bibnamefont{Scalapino}},
  \bibnamefont{and} \bibinfo{author}{\bibfnamefont{R.~T.}
  \bibnamefont{Scalettar}}, \bibinfo{journal}{Phys. Rev. B}
  \textbf{\bibinfo{volume}{45}}, \bibinfo{pages}{5577} (\bibinfo{year}{1992}),
  \urlprefix\url{https://link.aps.org/doi/10.1103/PhysRevB.45.5577}.

\bibitem[{\citenamefont{Terashima et~al.}(2014)\citenamefont{Terashima,
  Kikugawa, Kiswandhi, Choi, Brooks, Kasahara, Watashige, Ikeda, Shibauchi,
  Matsuda et~al.}}]{Terashima14}
\bibinfo{author}{\bibfnamefont{T.}~\bibnamefont{Terashima}},
  \bibinfo{author}{\bibfnamefont{N.}~\bibnamefont{Kikugawa}},
  \bibinfo{author}{\bibfnamefont{A.}~\bibnamefont{Kiswandhi}},
  \bibinfo{author}{\bibfnamefont{E.-S.} \bibnamefont{Choi}},
  \bibinfo{author}{\bibfnamefont{J.~S.} \bibnamefont{Brooks}},
  \bibinfo{author}{\bibfnamefont{S.}~\bibnamefont{Kasahara}},
  \bibinfo{author}{\bibfnamefont{T.}~\bibnamefont{Watashige}},
  \bibinfo{author}{\bibfnamefont{H.}~\bibnamefont{Ikeda}},
  \bibinfo{author}{\bibfnamefont{T.}~\bibnamefont{Shibauchi}},
  \bibinfo{author}{\bibfnamefont{Y.}~\bibnamefont{Matsuda}},
  \bibnamefont{et~al.}, \bibinfo{journal}{Phys. Rev. B}
  \textbf{\bibinfo{volume}{90}}, \bibinfo{pages}{144517}
  (\bibinfo{year}{2014}),
  \urlprefix\url{https://link.aps.org/doi/10.1103/PhysRevB.90.144517}.

\bibitem[{\citenamefont{L\"auchli et~al.}(2006)\citenamefont{L\"auchli, Schmid,
  and Trebst}}]{Trebst06}
\bibinfo{author}{\bibfnamefont{A.}~\bibnamefont{L\"auchli}},
  \bibinfo{author}{\bibfnamefont{G.}~\bibnamefont{Schmid}}, \bibnamefont{and}
  \bibinfo{author}{\bibfnamefont{S.}~\bibnamefont{Trebst}},
  \bibinfo{journal}{Phys. Rev. B} \textbf{\bibinfo{volume}{74}},
  \bibinfo{pages}{144426} (\bibinfo{year}{2006}),
  \urlprefix\url{http://link.aps.org/doi/10.1103/PhysRevB.74.144426}.

\bibitem[{\citenamefont{Manmana et~al.}(2011)\citenamefont{Manmana, L\"auchli,
  Essler, and Mila}}]{Manmana11}
\bibinfo{author}{\bibfnamefont{S.~R.} \bibnamefont{Manmana}},
  \bibinfo{author}{\bibfnamefont{A.~M.} \bibnamefont{L\"auchli}},
  \bibinfo{author}{\bibfnamefont{F.~H.~L.} \bibnamefont{Essler}},
  \bibnamefont{and} \bibinfo{author}{\bibfnamefont{F.}~\bibnamefont{Mila}},
  \bibinfo{journal}{Phys. Rev. B} \textbf{\bibinfo{volume}{83}},
  \bibinfo{pages}{184433} (\bibinfo{year}{2011}),
  \urlprefix\url{http://link.aps.org/doi/10.1103/PhysRevB.83.184433}.

\bibitem[{\citenamefont{Schmitt et~al.}(1998)\citenamefont{Schmitt, M\"utter,
  Karbach, Yu, and M\"uller}}]{Schmitt98}
\bibinfo{author}{\bibfnamefont{A.}~\bibnamefont{Schmitt}},
  \bibinfo{author}{\bibfnamefont{K.-H.} \bibnamefont{M\"utter}},
  \bibinfo{author}{\bibfnamefont{M.}~\bibnamefont{Karbach}},
  \bibinfo{author}{\bibfnamefont{Y.}~\bibnamefont{Yu}}, \bibnamefont{and}
  \bibinfo{author}{\bibfnamefont{G.}~\bibnamefont{M\"uller}},
  \bibinfo{journal}{Phys. Rev. B} \textbf{\bibinfo{volume}{58}},
  \bibinfo{pages}{5498} (\bibinfo{year}{1998}),
  \urlprefix\url{http://link.aps.org/doi/10.1103/PhysRevB.58.5498}.

\bibitem[{\citenamefont{Takhtajan}(1982)}]{Takhtajan82}
\bibinfo{author}{\bibfnamefont{L.~A.} \bibnamefont{Takhtajan}},
  \bibinfo{journal}{Phys. Lett.} \textbf{\bibinfo{volume}{A87}},
  \bibinfo{pages}{479} (\bibinfo{year}{1982}).

\bibitem[{\citenamefont{Babujian}(1982)}]{Babujian82}
\bibinfo{author}{\bibfnamefont{H.~M.} \bibnamefont{Babujian}},
  \bibinfo{journal}{Phys. Lett.} \textbf{\bibinfo{volume}{A90}},
  \bibinfo{pages}{479} (\bibinfo{year}{1982}).

\end{thebibliography}

\onecolumngrid

\section*{SM1: Schwinger boson mean field theory}

We show here that our ansatz state is a self-consistent solution of the $J_1$-$J_2$-$J_3$-$J_4$ spin model (see Fig.1a of main text for definition of $J$'s). On a bipartite lattice, it is convenient to perform a unitary transformation by defining $a_{im}\equiv b_{im}$ on A sublattice, and $a_{j\uparrow}\equiv b_{j\downarrow}$, $a_{j\downarrow}\equiv -b_{j\uparrow}$ on B sublattice. The valence bond operator is then brought to the simpler form $A^\dagger_{ij}= \sum_\sigma a^\dagger_{i\sigma}a^\dagger_{j\sigma} $. Modular a constant, the spin Hamiltonian 
${\cal H}_S=\sum_{ij}J_{ij}{\bm S}_i\cdot {\bm S}_j$
  can be written in terms of the valence bond operators as
   \begin{equation}
  {\cal H}_S=-\frac{1}{2}\sum_{ij}J_{ij}A^\dagger_{ij}A_{ij}.
   \end{equation}
    We then apply mean field theory to the bosonic Hamiltonian \cite{Arovas88}. Defining $Q_{ij}=J_{ij}\langle A_{ij}\rangle\equiv Q_\delta$, the quadratic part of the mean field Hamiltonian reads:  \begin{equation}
{\cal H}_S^{({\rm MF})}=\lambda \sum_{i\sigma} a^{\dagger}_{i\sigma}a_{i\sigma}+\frac{1}{2}\sum_{i\delta \sigma}Q_{\delta}\left( a^\dagger_{i\sigma}a^\dagger_{i+\delta, \sigma}+ a_{i\sigma}a_{i+\delta, \sigma}\right).
 \end{equation}
  For a given mean field ansatz, the mean field Hamiltonian can be diagonalized by the Bogoliubov transformation
\begin{equation}
 \alpha_{{\bm k}\sigma}=\cosh\theta_{\bm k}a_{{\bm k}\sigma}-\sinh\theta_{\bm k}a^\dagger_{-{\bm k}\sigma},
  \end{equation}
   with $\tanh \left( 2\theta_{\bm k} \right)=-Q\gamma_{\bm k}/\lambda$. Here $Q\gamma_{\bm k}$ denotes the Fourier transform of $Q_\delta$: $Q\gamma_{\bm k}=\sum_{\delta}Q_{\delta}e^{-i{\bm k}\cdot{\bm \delta}}$. 
 The resulting Hamiltonian reads 
 \begin{equation}
 {\cal H}_S^{({\rm MF})}=\sum_{{\bm k}\sigma}\omega_{\bm k}\left(\alpha^\dagger_{{\bm k}\sigma}\alpha_{{\bm k}\sigma}+\frac{1}{2}  \right),
  \end{equation}
with the dispersion $\omega_{\bm k}=\sqrt{\lambda^2-(Q\gamma_{\bm k})^2}$.  Integrating out the bosonic fields, one obtains the free energy 
  \begin{equation}
 F=\sum_{\delta}\frac{|Q_\delta|^2}{2J_{\delta}} -\frac{1}{2}(2S+1)\lambda+\frac{1}{\beta}\int \frac{d^2{\bm k}}{(2\pi)^2}\ln\left[2\sinh\left(\frac{1}{2}\beta\omega_{\bm k}\right)\right],
   \end{equation}
from which follow the self-consistency equations. In SBMFT, long-range order occurs through Bose-Einstein condensation (BEC) of the Schwinger bosons, and condensation gives rise to gapless spectrum due to the resulting Goldstone mode. A QSL state corresponds to a solution of the self-consistency equations with gapped spectrum, where there is no condensation of Schwinger bosons.

\begin{figure}[h]
\begin{centering}
\includegraphics[width=.8\textwidth]{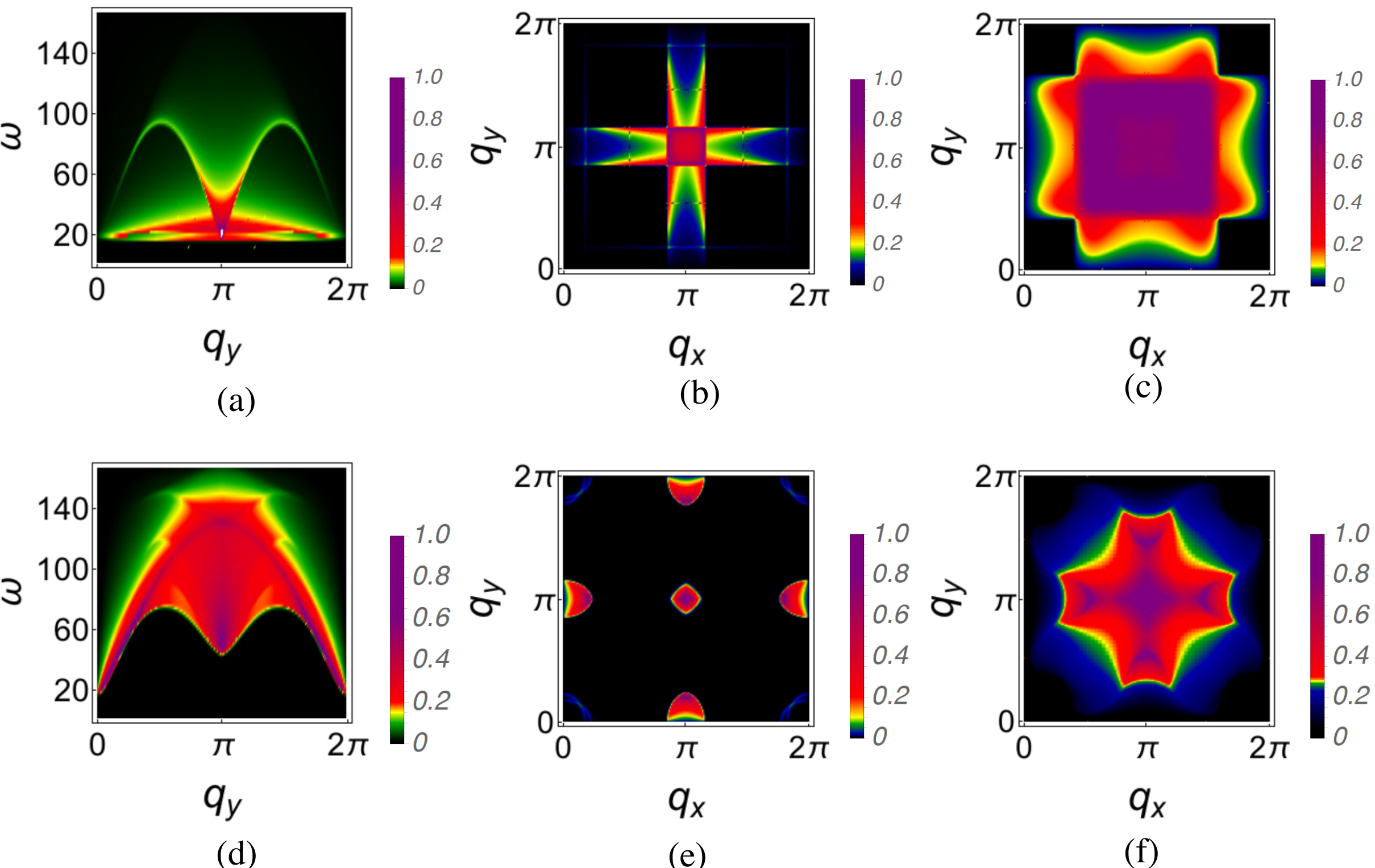}
\end{centering}
\caption{(Color online) Comparing the dynamic spin structure factor ${\cal S}(q_x, q_y, \omega)$ for decoupled 1d chain state (a, b, c) and quantum melted stripe state (d, e, f). (a, d): $q_x=\pi$; (b, e): $\omega=50$ meV; (c, f): $\omega=100$ meV. }
\label{FigSM:Chi}
\end{figure}

We start with decoupled 1d chains where only $Q_x\neq 0$, and 
 \begin{equation}
  Q\gamma_{\bm k}=2Q_x\cos k_x.
  \end{equation}
 Its free energy is
  \begin{equation}
 F^{(1)}=\frac{Q_x^2}{J_1}-\frac{1}{2}(2S+1)\lambda+\frac{1}{\beta}\int \frac{d^2{\bm k}}{(2\pi)^2}\ln\left[2\sinh\left(\frac{1}{2}\beta\omega_{\bm k}\right)\right],
   \end{equation}
   and the self-consistency equations are 
 \begin{eqnarray}
 S+\frac{1}{2}&=& \int \frac{d^2{\bm k}}{(2\pi)^2}\frac{\lambda}{2\omega_{\bm k}}, 
 \\
\frac{Q_x}{J_1}&=& \int \frac{d^2{\bm k}}{(2\pi)^2}\frac{Q\gamma_{\bm k}}{2\omega_{\bm k}}\cos k_x.
\end{eqnarray}  
This ansatz state is basically determined by the dimensionless parameters: $Q_x/\lambda$. We have plotted the spin structure factor taking $Q_x/\lambda=0.498$ (see Fig.~\ref{FigSM:Chi}a,b,c). Here $S=0.677$.
  
 We consider then our ansatz state, namely the quantum melted stripe state, where $Q_x\neq 0$, $Q_{x+y}\neq 0$, $Q_{x+2y}\neq 0$, and
     \begin{equation}
  Q\gamma_{\bm k}=2Q_x\cos k_x+4Q_{x+y}\cos k_x \cos k_y  + 4Q_{x+2y}\cos k_x \cos(2k_y).
  \end{equation}
Its free energy is
  \begin{equation}
 F^{(2)}=\frac{Q_x^2}{J_1}+ \frac{2Q_{x+y}^2}{J_2}+\frac{2Q_{x+2y}^2}{J_4}-\frac{1}{2}(2S+1)\lambda+\frac{1}{\beta}\int \frac{d^2{\bm k}}{(2\pi)^2}\ln\left[2\sinh\left(\frac{1}{2}\beta\omega_{\bm k}\right)\right],
   \end{equation}
   and the self-consistency equations are 
 \begin{eqnarray}
 S+\frac{1}{2}&=& \int \frac{d^2{\bm k}}{(2\pi)^2}\frac{\lambda}{2\omega_{\bm k}}, 
 \\
\frac{Q_x}{J_1}&=& \int \frac{d^2{\bm k}}{(2\pi)^2}\frac{Q\gamma_{\bm k}}{2\omega_{\bm k}}\cos k_x, 
\\
\frac{Q_{x+y}}{J_2}&=& \int \frac{d^2{\bm k}}{(2\pi)^2}\frac{Q\gamma_{\bm k}}{2\omega_{\bm k}}\cos k_x\cos k_y ,
\\
\frac{Q_{x+2y}}{J_4}&=& \int \frac{d^2{\bm k}}{(2\pi)^2}\frac{Q\gamma_{\bm k}}{2\omega_{\bm k}}\cos k_x\cos(2k_y).
\end{eqnarray}  
This ansatz state is basically determined by the three dimensionless parameters: $Q_x/\lambda$, $Q_{x+y}/Q_x$ and $Q_{x+2y}/Q_x$. We have plotted the spin structure factor taking $Q_x/\lambda=0.398$, $Q_{x+y}/Q_x=0.025$ and $Q_{x+2y}/Q_x=0.1$ (see Fig.~\ref{FigSM:Chi}d,e,f). Here $S=0.153$, $J_2/J_1=0.904$, $J_4/J_1=0.975$.

In addition, we find that the quantum melted stripe state has lower energy than the decoupled 1d chain state. We set $J_1=1$. For $S=0.25$,  $J_2=0.88$, $J_4=0.944$, we obtain $F^{(1)}=-0.175$,  $F^{(2)}=-0.179$. For $S=0.26$, $J_2=0.864$, $J_4=0.925$, we obtain $F^{(1)}=-0.188$, $F^{(2)}=-0.192$. For $S=0.3$, $J_2=0.869$, $J_4=0.935$, we obtain $F^{(1)}=-0.224$, $F^{(2)}=-0.227$. 

\section*{SM2: Itinerant part: three orbital model}

For the itinerant part of the system, what matters for pairing are the low energy electronic states around the Fermi pockets. We take a phenomenological approach to expand the dispersion around the Fermi pockets. Since it is known experimentally that the spectral weight of the low energy states arises mainly from $d_{yz}$, $d_{zx}$, $d_{xy}$ orbitals, we consider a band structure involving these three orbitals. Such orbital-projected band models have been studied in \cite{Vafek13, Fernandes16}. Consider first the Fermi pocket near the $\Gamma$ point, a single hole pocket has been detected, with the $d_{yz}$ and $d_{zx}$ orbitals dominating the spectral weight. We introduce a spinor $\psi^T_{\gamma, {\bm k}}=(c_{yz, {\bm k}\uparrow}, -c_{zx, {\bm k}\uparrow}, c_{yz, {\bm k}\downarrow}, -c_{zx, {\bm k}\downarrow})$, and the kinetic energy part of the Hamiltonian is of the form 
\begin{equation}
{\cal H}_{0,\Gamma} = \sum_{\bm k} \psi^\dagger_{\gamma, {\bm k}} h_{\Gamma}({\bm k})\psi_{\Gamma, {\bm k}}.
\end{equation}
The Hamiltonian includes the on-site energy, intra-orbital hopping, inter-orbital hopping. To get the right orbital splitting, we include also the difference in the on-site energy for the two orbitals reflecting nematicity, and the spin-orbit coupling \cite{Vafek13}. The result reads
\begin{equation}
h_{\Gamma}({\bm k})=\left(\varepsilon_\Gamma+\frac{k^2}{2m_\Gamma}\right) \tau^0\otimes \sigma^0 + \left[\delta\varepsilon_\Gamma +b k^2\cos (2\theta_k)\right] \tau^3\otimes \sigma^0 + ck^2\sin (2\theta_k) \tau^1\otimes \sigma^0 + \lambda \tau^2\otimes \sigma^3,
\end{equation}
where $\tau$ and $\sigma$ are Pauli matrices in orbital and spin space respectively, and ${\bm k}=(k_x, k_y)=k(\cos\theta_k, \sin\theta_k)$.

For the electron pocket near $(\pi, 0)$, the $d_{yz}$ and $d_{xy}$ orbitals dominate the spectral weight.
We introduce a spinor $\psi^T_{X, {\bm k}}=(c_{yz, {\bm k}\uparrow}, c_{xy, {\bm k}\uparrow}, c_{yz, {\bm k}\downarrow}, c_{xy, {\bm k}\downarrow})$, and the kinetic energy part of the Hamiltonian is of the form 
\begin{equation}
{\cal H}_{0,X} = \sum_{\bm k} \psi^\dagger_{X, {\bm k}} h_X({\bm k})\psi_{X, {\bm k}}, 
\end{equation}
with 
 \begin{equation}
h_X({\bm k})=\left[\varepsilon_1 +\frac{k^2}{2m_1}-a_1 k^2\cos(2\theta_k)\right] \frac{\tau^0+\tau^3}{2}\otimes \sigma^0 +\left[\varepsilon_3 +\frac{k^2}{2m_3}-a_3 k^2\cos(2\theta_k)\right] \frac{\tau^0-\tau^3}{2}\otimes \sigma^0+2vk\sin\theta_k \tau^2\otimes \sigma^0.
\end{equation}
Here ${\bm k}$ is measured from $(\pi, 0)$.

For the electron pocket near $(0, \pi)$, the $d_{zx}$ and $d_{xy}$ orbitals dominate the spectral weight. We introduce a spinor $\psi^T_{Y, {\bm k}}=(c_{zx, {\bm k}\uparrow}, c_{xy, {\bm k}\uparrow}, c_{zx, {\bm k}\downarrow}, c_{xy, {\bm k}\downarrow})$, and the kinetic energy part of the Hamiltonian is of the form 
\begin{equation}
{\cal H}_{0,Y} = \sum_{\bm k} \psi^\dagger_{Y, {\bm k}} h_Y({\bm k})\psi_{Y, {\bm k}}, 
\end{equation}
with 
 \begin{equation}
h_Y({\bm k})=\left[\varepsilon_1 +\frac{k^2}{2m_1}+a_1 k^2\cos(2\theta_k)\right] \frac{\tau^0+\tau^3}{2}\otimes \sigma^0 +\left[\varepsilon_3 +\frac{k^2}{2m_3}+a_3 k^2\cos(2\theta_k)\right] \frac{\tau^0-\tau^3}{2}\otimes \sigma^0+2vk\cos\theta_k \tau^2\otimes \sigma^0.
\end{equation}
Here ${\bm k}$ is measured from $(0, \pi)$.

With a proper choice of the parameters, we can obtain a single hole pocket around $\Gamma$, a single electron pocket around $(\pi, 0)$, and a single electron pocket around $(0, \pi)$ as shown in Fig.4a of main text. The corresponding parameters are: $\varepsilon_\Gamma =14$, $\delta\varepsilon_\Gamma = 11$, $\frac{1}{2m_\Gamma}=-350$, $b=-70$, $c=120$, $\lambda = 9$, $\varepsilon_1=-20$, $\varepsilon_3=-60$ $\frac{1}{2m_1}=75$, $\frac{1}{2m_3}=160$, $a_1=100$, $a_3=-120$, $v=-60$. Note that the band structure employed in our paper is simplified in order to use a closed form for hamiltonian that allows us to carry out the study of superconductivity semi-analytically. Although our band structure misses  quantitative details such as severe mismatch in the pocket sizes between two electron pockets (see SM 4 for further discussion), such details will not impact qualitative conclusions of the paper. 

\section*{SM3: How itinerant electrons affect local moments: Landau damping}

\begin{figure}[h]
\begin{centering}
\includegraphics[width=.8\textwidth]{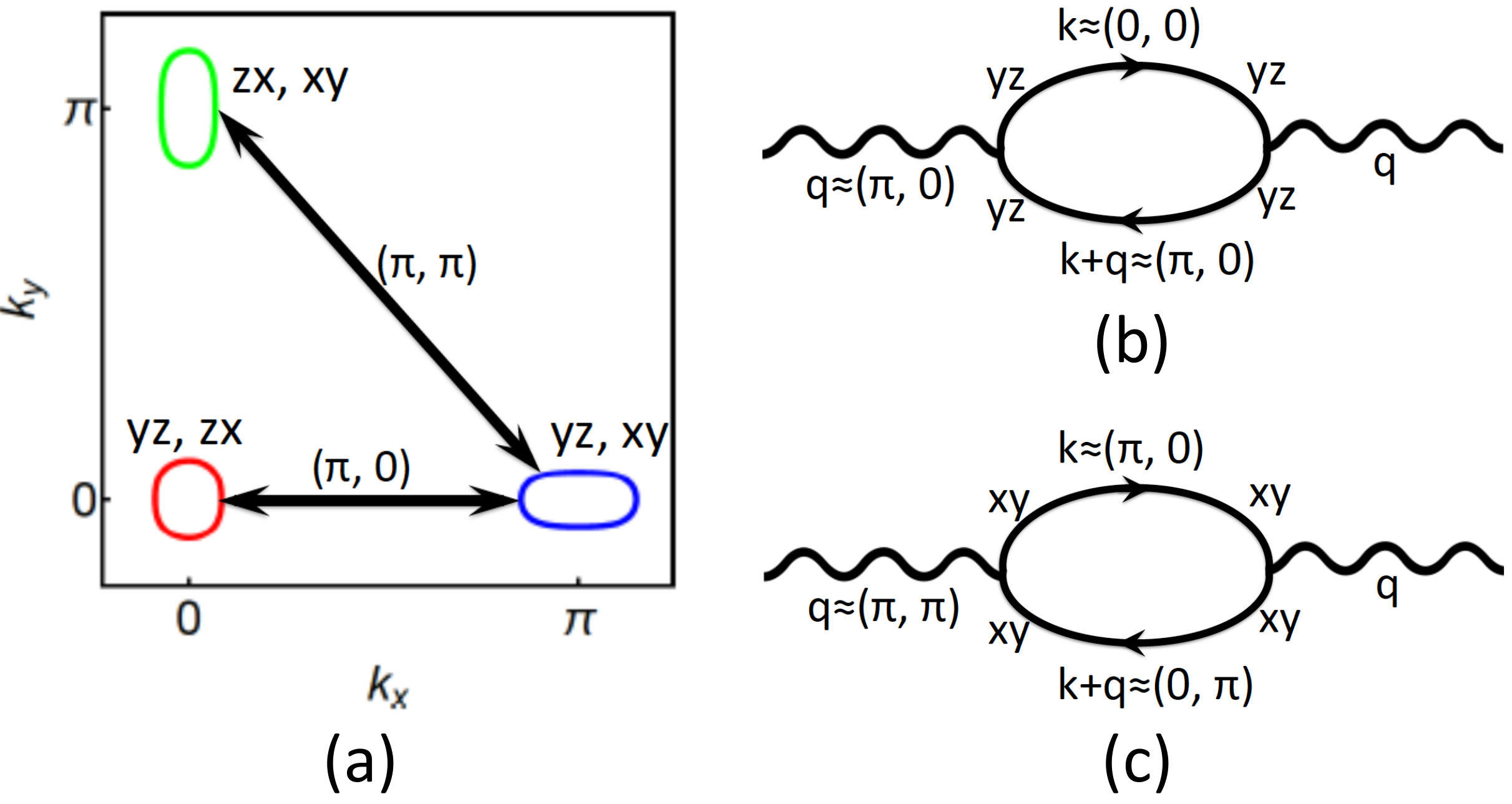}
\end{centering}
\caption{(Color online) (a) Fermi pockets (with orbital contents labeled) and the resulting scattering processes that contribute to local moment self-energy. (b, c) The corresponding Feynman diagrams. }
\label{FigSM:damping}
\end{figure}

Coupling to itinerant electrons generates a self-energy for the local moment propagator, giving rise to Landau damping. Since the Fermi pockets are small in size, and located near $\Gamma$- and M-points, the induced self-energy will be predominantly near ${\bm q}=(\pi, \pi)$, $(\pi, 0)$ and $(0, \pi)$ (see Fig.\ref{FigSM:damping}). We expect the neutron spectrum near these points to be smeared. For ${\bm q}\sim (\pi, 0)$, the self-energy is predominantly from $d_{yz}$ orbitals, 
 \begin{equation}
 {\cal D}^{(\pi, 0)}({\bm q}, \Omega) \sim \sum_{{\bm k},\omega}J_{yz}^2 G^{(yz)}({\bm k}, \omega) G^{(yz)}({\bm k}+{\bm q}, \omega+ \Omega),
  \end{equation}
with fermion Green's function $G({\bm k}, \omega)$. Here ${\bm k}$ is at the $\Gamma$ pocket, and ${\bm k}+{\bm q}$ at the $M_x$ pocket. For ${\bm q}\sim (\pi, \pi)$, the self-energy is predominantly from $d_{xy}$ orbitals, 
 \begin{equation}
 {\cal D}^{(\pi, \pi)}({\bm q}, \Omega) \sim \sum_{{\bm k},\omega}J_{xy}^2 G^{(xy)}({\bm k}, \omega) G^{(xy)}({\bm k}+{\bm q}, \omega+ \Omega),
  \end{equation}
where ${\bm k}$ and ${\bm k}+{\bm q}$ are at the two $M$-pockets. With the suppression of $J_{xy}$, we expect the Landau damping effect to be weaker near ${\bm q}=(\pi, \pi)$.

We can estimate the strength of the Hund's coupling from the resulting superconducting transition temperature: $T_c\sim E_F e^{-1/\lambda}$, with the dimensionless coupling $\lambda= N_0 V\sim J_H^2/(E_FJ_{\rm ex})$. The energy scales involved are: (1) Fermi energy $E_F\sim 10$ meV \cite{Kasahara14}, (2) the exchange interaction $J_{\rm ex}\sim 100$ meV \cite{Mazin15}, and (3) $T_c\sim 1$ meV. From these, we obtain Hund's coupling $J_H\sim 20$ meV, which is much smaller than the exchange interaction $J_{\rm ex}$. Hence we expect coupling to itinerant electrons will not significantly modify the local moment spin susceptibility.

\section*{SM4: How local moments affect itinerant electrons: nematicity and pairing}

The dynamic spin fluctuations in the QSL affect the itinerant electrons. Since the spins have a gapped spectrum, we can integrate them out to obtain an effective interaction for the itinerant electrons. The induced action reads
\begin{equation}
{\cal S}_{\rm int} =-\frac{1}{2}\int_0^\beta d\tau\sum_{\alpha, \alpha'} J_{\alpha}J_{\alpha'}\chi_{ij}(\tau){\bm s}_{i\alpha}(\tau)\cdot{\bm s}_{j\alpha'}(0),
\end{equation}
with the itinerant electron spin density ${\bm s}_{i\alpha}\equiv \sum_{\mu\nu}c^\dagger_{i\alpha\mu}{\bm \sigma}_{\mu\nu}c_{i\alpha\nu}$, and the local moment spin correlation function $\chi_{ij}(\tau)\equiv\langle T_{\tau}S_i^a(\tau)S_j^a(0) \rangle$. The induced interaction is highly anisotropic, and the dominant interaction term is the nearest-neighbor interaction (say along the $x$-direction): $J_H^2\chi c^\dagger_{{\bm r}\alpha}\sigma^a_{\alpha\beta}c_{{\bm r}\beta}c^\dagger_{{\bm r}+{\hat x},\alpha'}\sigma^a_{\alpha'\beta'}c_{{\bm r}+{\hat x},\beta'}$. This interaction results in a phase transition to a nematic state with order parameter $\langle c^\dagger_{{\bm r}+{\hat x},\alpha}c_{{\bm r},\alpha} \rangle \neq 0$, or more generally, $\varphi_c\equiv  \langle c^\dagger_{{\bm r}+{\hat x},\alpha}c_{{\bm r},\alpha} -c^\dagger_{{\bm r}+{\hat y},\alpha}c_{{\bm r},\alpha} \rangle \neq 0$.

\begin{figure}[h]
\begin{centering}
\includegraphics[width=.4\textwidth]{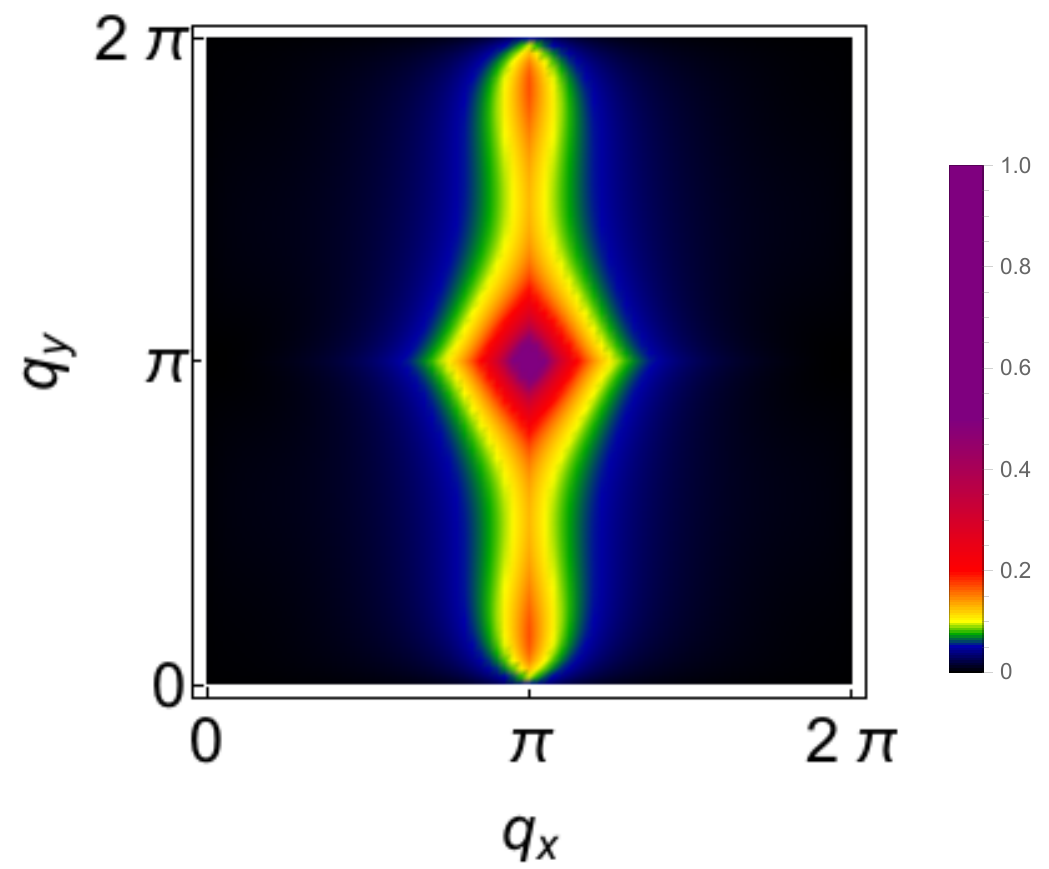}
\end{centering}
\caption{(Color online) The spin susceptibility $\chi({\bm q})\equiv \chi({\bm q}, \Omega_n=0)$ in a single nematic domains. }
\label{FigSM:chi}
\end{figure}

Furthermore, the induced interaction leads to pairing among the itinerant electrons.
Since the spin fluctuations are antiferromagnetic, one expects pairing in the spin singlet channel. We then mean field decompose the induced interaction into spin singlet pairing channel with the corresponding pair operator $h^\dagger_{\alpha\alpha'}({\bm k})=\frac{1}{\sqrt{2}}\left(c^\dagger_{{\bm k}\alpha\uparrow}c^\dagger_{-{\bm k}\alpha'\downarrow} -c^\dagger_{{\bm k}\alpha\downarrow}c^\dagger_{-{\bm k}\alpha'\uparrow}\right)$. Due to the special form of spin susceptibility and band structure in FeSe, the pairing problem is largely simplified. The spin fluctuations enter the pairing problem through the spin susceptibility $\chi({\bm q})\equiv \chi({\bm q}, \Omega_n=0)$, which can be obtained from the dynamic spin structure factor via $\chi({\bm q})=-\int d\omega {\cal S}_{{\bm q}, \omega}/\omega$. The special form of $\chi({\bm q})$ in FeSe (see Fig.~\ref{FigSM:chi}) results in only inter-band pairing correlation among the three Fermi pockets. Furthermore, since Hund's coupling is diagonal in orbital space, there are only pairing correlations between the same orbitals: in orbital basis, the pairing interaction is of the form ${\cal H}_{\rm pair} \sim J_\alpha^2 \chi({\bm k}-{\bm k}')h^\dagger_{\alpha\alpha}({\bm k})h_{\alpha\alpha}({\bm k}')$. 

Pairing occurs near the Fermi surface, which is naturally expressed in the band basis. We then transform from the orbital basis to the band basis: $c^\dagger_{{\bm k}\alpha\mu}=\sum_a \eta^*_{\alpha a \mu}({\bm k})d^\dagger_{{\bm k}a\mu}$ with band index $a$. Note that since the spin-orbit coupling here is in the $\sigma^3$ channel, different spins do not mix. The pair operator in the band basis is $h^\dagger_a({\bm k})=\frac{1}{\sqrt{2}}\left(d^\dagger_{{\bm k}a\uparrow}d^\dagger_{-{\bm k}a\downarrow} -d^\dagger_{{\bm k}a\downarrow}d^\dagger_{-{\bm k}a\uparrow}\right)$. Omitting the frequency dependence, the pairing Hamiltonian is of the form 
\begin{equation}
{\cal H}_{\rm pair} =\sum_{{\bm k}{\bm k}'ab}\Gamma_{ab}({\bm k}, {\bm k}')h^\dagger_a({\bm k})h_b({\bm k}'),
\end{equation}
 with the projected pairing interaction $ \Gamma_{ab}({\bm k}, {\bm k}')=\frac{1}{2}\sum_{\alpha}J_{\alpha}^2\chi({\bm k}-{\bm k}') M^*_{\alpha a}({\bm k}) M_{\alpha b}({\bm k}')$. The orbital content is encoded in the form factor $M_{\alpha a}({\bm k})= \eta_{\alpha a \uparrow}({\bm k}) \eta_{\alpha a \downarrow}(-{\bm k})$.
  The gap function is then defined as $\Delta_a({\bm k}) =\sum_{{\bm k}'b}\Gamma_{ab}({\bm k}, {\bm k}')\langle h_b({\bm k}')\rangle$. The gap symmetry function $g_i({\bm k})\propto \Delta_a({\bm k})$ on the Fermi surface is determined by the eigen equation
\begin{equation}
-\sum_j\oint_{{\rm FS}_j} \frac{d{\bm k}'_{\parallel}}{2\pi v_F({\bm k}')} \Gamma_{ij}({\bm k}, {\bm k}')g_j({\bm k}') =\lambda g_i({\bm k}),
\label{Eq:eigen}
\end{equation}
where ${\bm k}_{\parallel}$ denotes momentum along the Fermi surface ${\rm FS}_j$, and $v_F({\bm k})=|\nabla E_a({\bm k})|$ represents the Fermi velocity. We can then solve the above eigen equation to find the leading eigenvalue and the corresponding eigenvector, which determines the resulting gap structure within a single nematic domain. The inputs are (1) itinerant electron band structure encoded in $\epsilon_{\alpha\beta}^{\mu\nu}({\bm k})$ (2) local moment spin susceptibility $\chi({\bm q})$, and (3) Hund's couplings $J_\alpha$.

We show here more concretely how $(\pi, \pi)$ spin fluctuation mediated pairing enhances the superconducting $T_c$.
We first estimate the value of $\lambda$ in bulk FeSe from the observed $T_c$. Since Fermi energy is small compared to spin fluctuation scale (so called antiadiabatic limit), Fermi energy acts as cutoff in the $T_c$ equation: $T_c\sim E_F e^{-1/\lambda}$ \cite{Kirzhnits1973, Marel14}. With $T_c\sim 8$ K, $E_F\sim 10$ meV \cite{Kasahara14}, we obtain $\lambda\sim 0.37$. In bulk FeSe, pairing occurs predominantly among $d_{yz}$ orbitals as mediated by $(\pi, 0)$ spin fluctuations, while $(\pi, \pi)$ spin fluctuation mediated pairing among $d_{xy}$ orbitals is largely suppressed. This corresponds to taking $(J_{xy}, J_{zx}, J_{yz}) \sim (0, 1, 1)$. (Note that due to the near absence of $(0, \pi)$ spin fluctuations, pairing among $d_{zx}$ orbitals is suppressed for any coupling. So we just set $J_{zx} =1$.) When pairing is predominantly among $d_{xy}$ orbitals, we have $(J_{xy}, J_{zx}, J_{yz}) \sim (1, 1, 0)$. We have obtained the resulting eigenvalue  $\lambda'=2.98\lambda$, which gives $T_c\sim 47$ K. Hence $(\pi, \pi)$ spin fluctuation mediated pairing is indeed able to account for the much higher $T_c$ in heavily doped FeSe ($T_c\sim 48$ K \cite{Miyata15}), and a large part of the $T_c$ increase in monolayer FeSe ($T_c\sim 50-64$ K \cite{ZXShen14}).  

\subsection*{Neutron resonance}

A characteristic feature of our gap function is that it has different signs at different Fermi pockets. Such a sign-changing gap function can give rise to resonances in the neutron spectrum. However the intensity of the resonances depends on the details of the band structure and the superconducting gap function. The resonance comes from the itinerant electron spin susceptibility, which contains a term of the form\cite{Bulut92}
\begin{equation}
\chi''({\bm q}, \omega)\sim \sum_{\bm k}\left( 1-\frac{\Delta_{\bm k}\Delta_{{\bm k}+{\bm q}}}{E_{\bm k}E_{{\bm k}+{\bm q}} }\right)
\left[ 1-f(E_{\bm k})-f(E_{{\bm k}+{\bm q}}) \right]\delta(\omega-E_{\bm k}-E_{{\bm k}+{\bm q}} ).
\end{equation}
The gap function changes sign between the $\Gamma$ pocket and $(\pi, 0)$ pocket, and between the $(\pi, 0)$ pocket and $(0, \pi)$ pocket. Since the sizes of the Fermi pockets are small, one expects the  resonances to be localized in momentum space around ${\bm q}=(\pi, 0)$ and ${\bm q}=(\pi, \pi)$.

\begin{figure}[t]
\begin{centering}
\subfigure[]{
\includegraphics[width=0.35\linewidth]{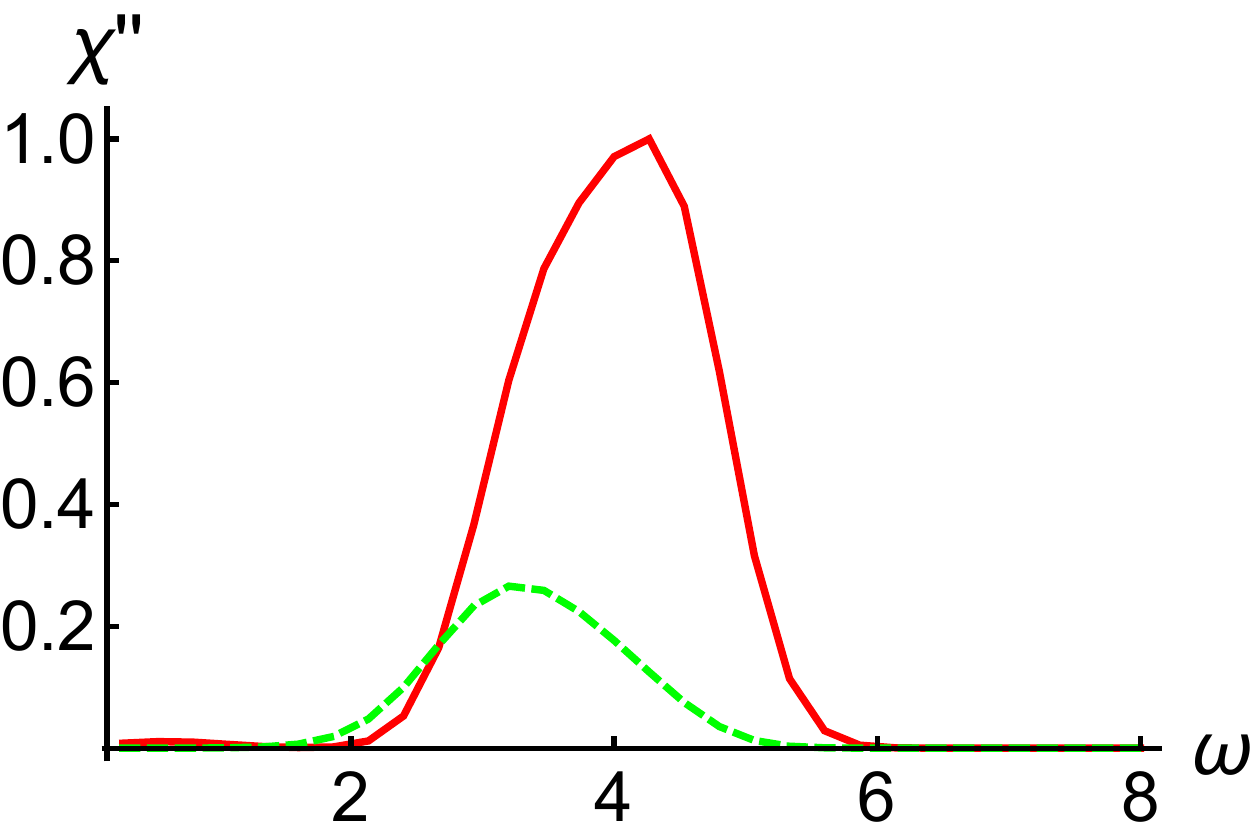}~~~ }
\subfigure[]{
\includegraphics[width=0.35\linewidth]{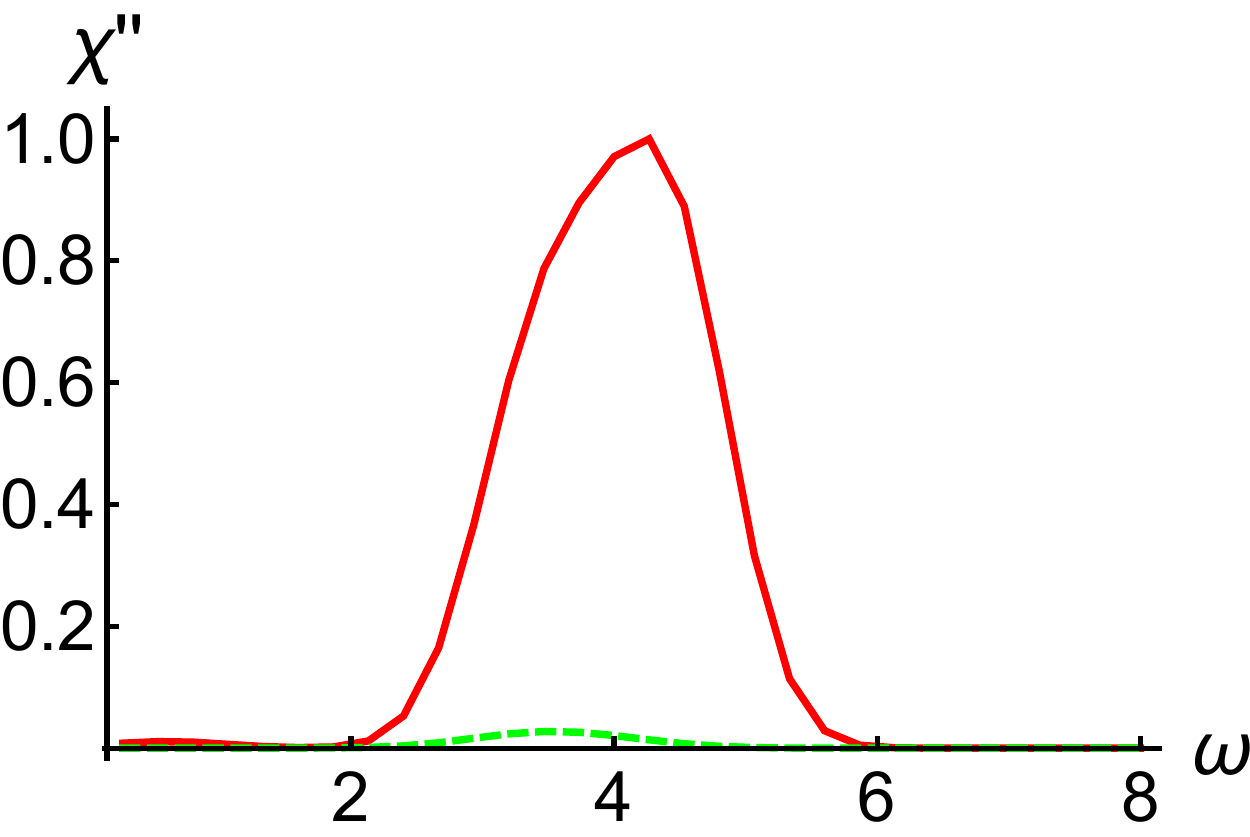} }
\end{centering}
\caption{(Color online) The low frequency neutron intensity at $(\pi, 0)$ (red solid line) and $(\pi, \pi)$ (green dashed line) resulting from the superconducting gap as shown in Fig.4(c) of main text for (a): the simplified band structure used in this paper, and (b): the band structure reflecting mismatch in the Fermi pocket sizes as detected by quantum oscillation. }
\label{FigSM:neutron}
\end{figure}

Due to lack of full knowledge of the gap function in the whole Brillouin zone, we approximate the neutron intensity at ${\bm q}=(\pi, 0)$ and ${\bm q}=(\pi, \pi)$ by summing over the corresponding Fermi surfaces. Furthermore we penalize the resulting term by an exponential factor depending on the difference between the momentum transfer and ${\bm q}$, i.e. we replace $\sum_{\bm k}$ by $\oint_{\bm k}\oint_{{\bm k}'}e^{-|{\bm k}-{\bm k}'-{\bm q}|^2/q_0^2}$, where the integrals are restricted to the Fermi surfaces, and $q_0$ is a parameter that basically measures the range of blurring in momentum space. Using the simplified band structure employed in the paper and specified in SM3 (Fig.4(a) of main text), and the resulting gap function with $J_{xy}/J_{yz}=0.4$ (see Fig.4(c) of main text), we obtain the neutron intensity as shown in Fig.\ref{FigSM:neutron}(a) (with $q_0=0.2$). One can see that the resonance at ${\bm q}=(\pi, \pi)$ is suppressed due to worse nesting of the corresponding Fermi surface and gap functions. 

Actually one expects further suppression of $(\pi, \pi)$ resonance with more realistic band structures.  In particular, it has been found in quantum oscillation measurements \cite{Terashima14} that the Fermi pocket at $(0, \pi)$ (with $k_F\simeq 0.13{\rm \AA}^{-1}$) is much larger than the Fermi pocket at $(\pi, 0)$ (with $k_F\simeq 0.043{\rm \AA}^{-1}$). Indeed a simple check by enlarging the Fermi pocket at $(0, \pi)$ by a factor of two while keeping the rest of the band structure and gap functions fixed suppreses the resonance at ${\bm q}=(\pi, \pi)$ to be almost vanishing (see Fig.\ref{FigSM:neutron}(b)).

\section*{SM5: AKLT chains or fractionalized spin liquid in FeSe?}
The theory of FeSe presented in the main manuscript consisted of coupling a nematic quantum spin liquid of local moments described by Schwinger boson mean field theory to itinerant electrons. In the absence of this coupling, the local moments would behave like they do in insulators where it is known \cite{Read90} that this Schwinger boson mean field theory is unstable by confinement and forms chains of AKLT states. This confined state is presumably related to the AKLT chain states studied in Ref. \onlinecite{FWang15} as a potential theory of the magnetism in FeSe. 

Here we attempt to find an experimental signature that could distinguish the deconfined state studied in the main manuscript and an AKLT chain state in a future experiment. Our approach will be to make use of known results from one dimensional physics. If we decouple chains in the AKLT chain state, we can use numerical results on the well studied one dimensional model to determine signatures in neutron scattering for such a state. Then after determining experimentally relevant features of this state we will step back and assess the implication of these results for the broader question of the distinction between the AKLT chain state and nematic quantum spin liquid states.

\begin{figure}[t]
\begin{centering}
\includegraphics[width=.4\textwidth]{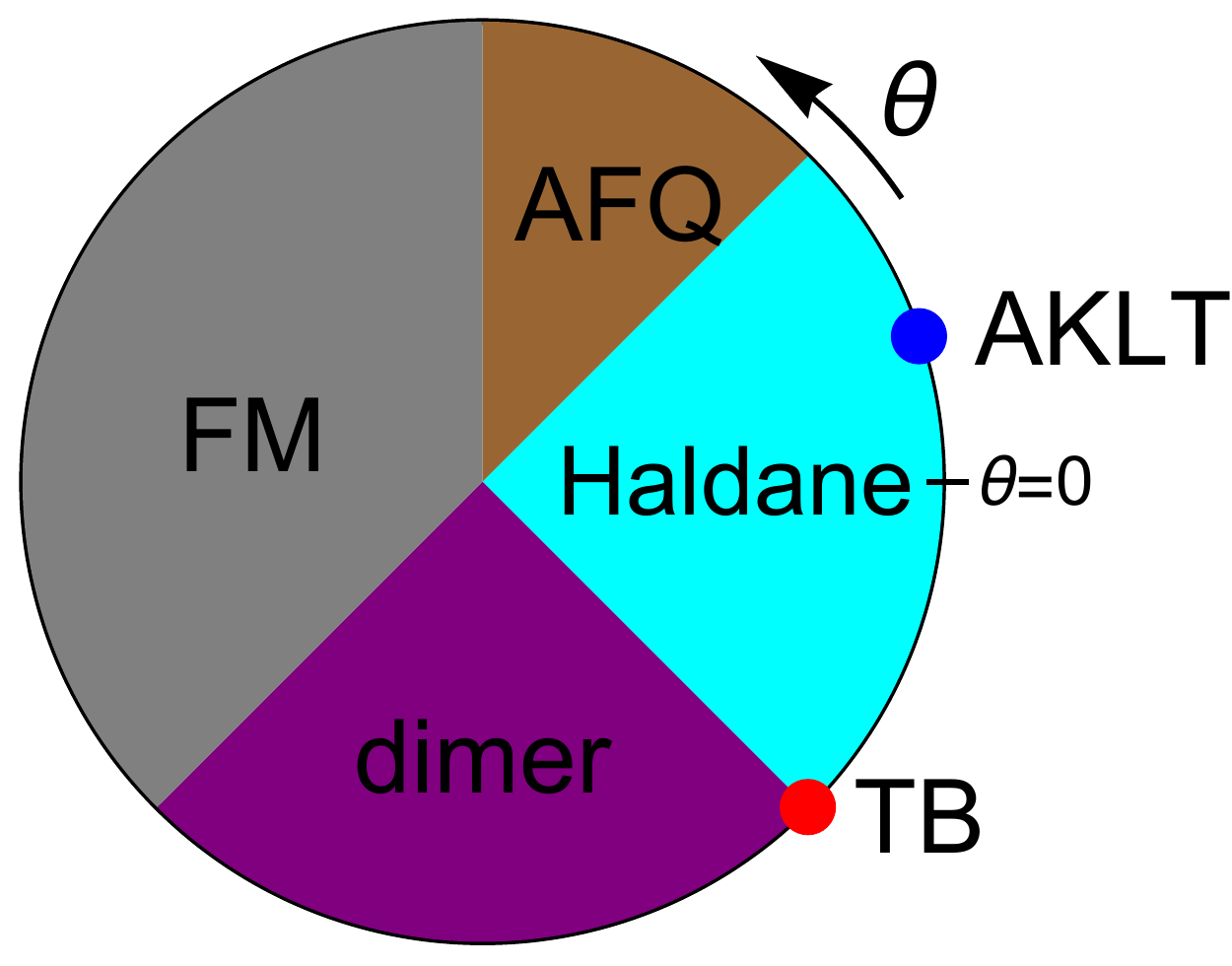}
\end{centering}
\caption{(Color online) Phase diagram of 1d $S=1$ spin chains with bilinear and biquadratic interactions \cite{Trebst06, Manmana11}. FM stands for ferromagnetic phase, AFQ for antiferroquadrupolar phase, AKLT for Affleck-Lieb-Kennedy-Tasaki model, TB for Takhtajan-Babujian model.}
\label{Fig:1d-chain}
\end{figure}

\begin{figure}[t]
\begin{centering}
\includegraphics[width=.4\textwidth]{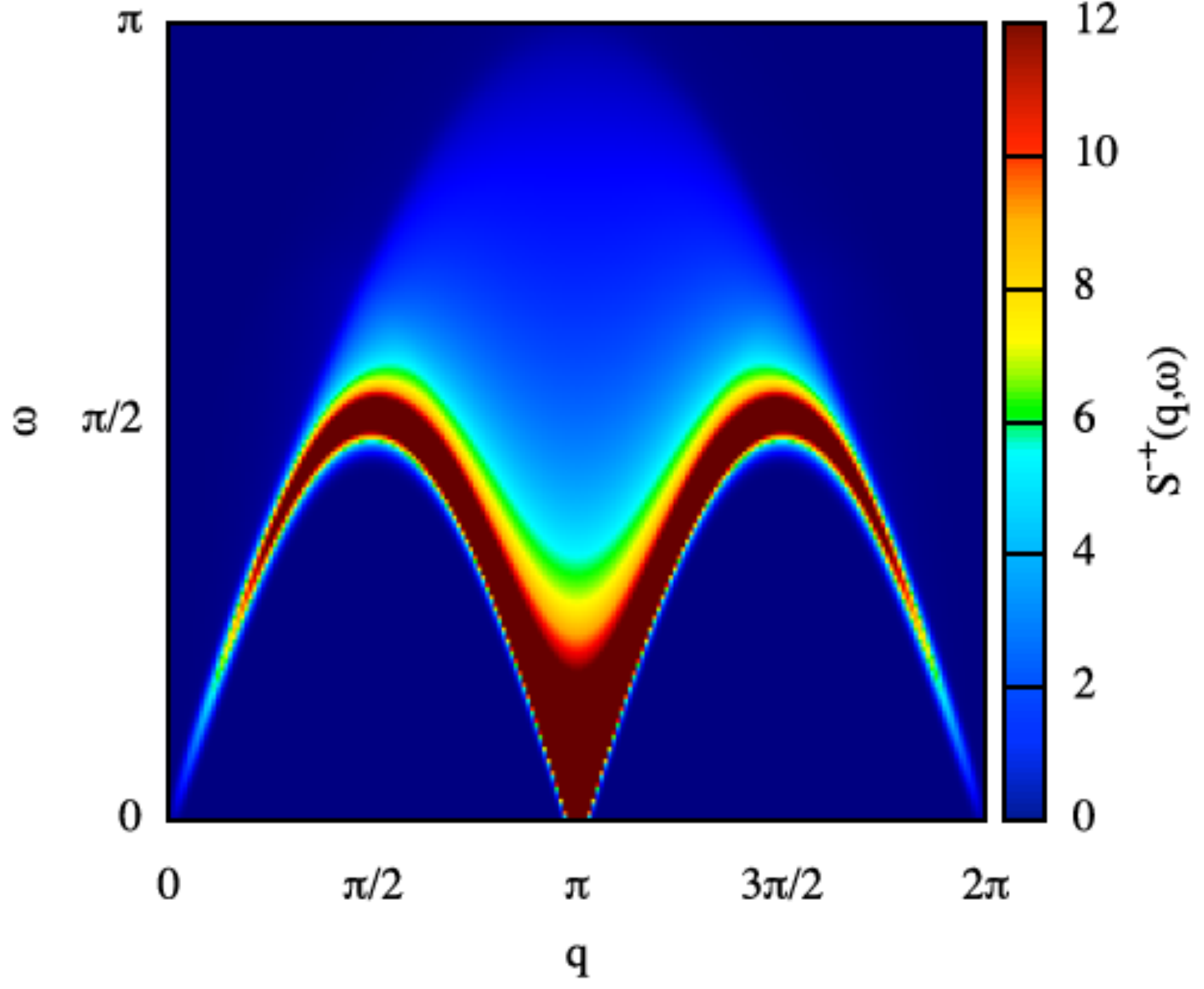}
\end{centering}
\caption{(Color online) Dynamic spin structure factor of Takhtajan-Babujian chain as obtained in Bethe Ansatz \cite{Caux14}. }
\label{Fig:TB-chain}
\end{figure}

\begin{figure}[t]
\begin{centering}
\includegraphics[width=.9\textwidth]{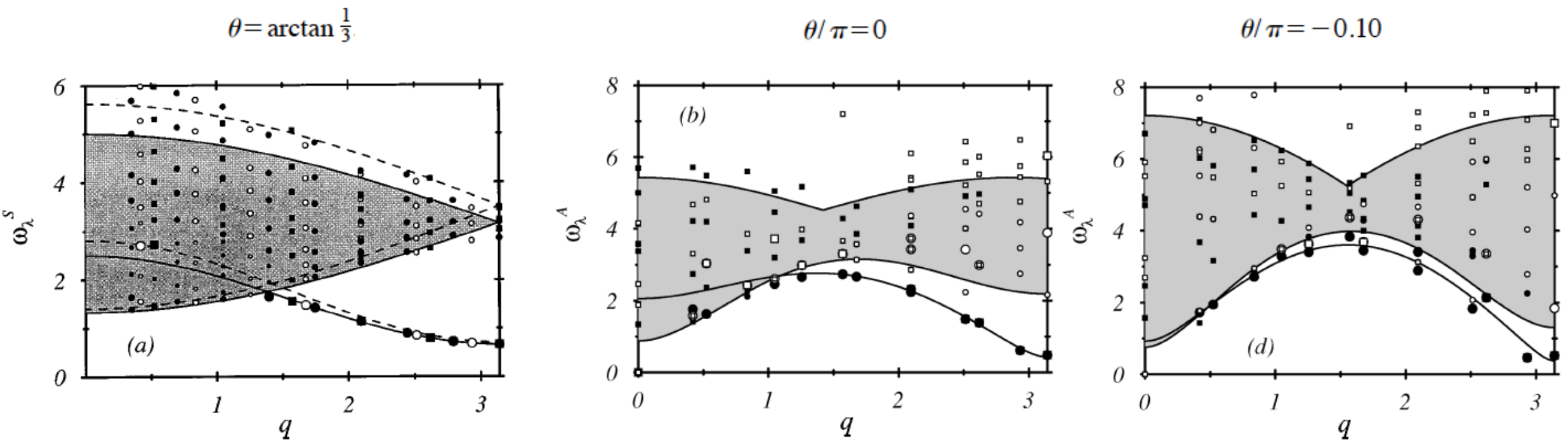}
\end{centering}
\caption{(Color online) Dynamic spin structure factor of 1d $S=1$ spin chains with the corresponding $\theta$ values as obtained in \cite{Schmitt98}. }
\label{Fig:BLBQ-chain}
\end{figure}

In one dimension, the $S=1$ spin chain with bilinear and biquadratic interactions has a variety of phases including the AKLT state and a variety of phase transitions. The Hamiltonian is
\begin{equation}
  {\cal H}_S=J\sum_{i}\left[\cos\theta \left({\bm S}_i\cdot {\bm S}_{i+1} \right) +\sin\theta \left({\bm S}_i\cdot {\bm S}_{i+1} \right)^2  \right],
   \end{equation}
with $-\pi\leq\theta\leq \pi$. As shown in Fig.\ref{Fig:1d-chain}, its phase diagram contains (see \cite{Trebst06,Manmana11} and references therein): (1) a ferromagnetic phase for $-\pi<\theta<-3\pi/4$ and $\pi/2<\theta\leq \pi$, (2) a dimerized phase for $-3\pi/4<\theta<-\pi/4$, (3) a gapped and topologically ordered Haldane phase for $-\pi/4<\theta<\pi/4$ (the AKLT state corresponds to $\theta=\arctan\frac{1}{3}\simeq 0.1024\pi$) and (4) a gapless phase with antiferroquadrupolar (AFQ) correlations for $\pi/4<\theta<\pi/2$ \cite{Trebst06}. The dimerized phase and the Haldane phase are separated by a critical point, the Takhtajan-Babujian (TB) point \cite{Takhtajan82, Babujian82}. At the TB point, the system possesses gapless spinon excitations. The spinon continuum is manifest in the dynamical spin structure factor (Fig.\ref{Fig:TB-chain}) as obtained using algebraic Bethe ansatz-based method \cite{Caux14}. As one moves away from the critical point into the Haldane phase, the spinons get confined, and the elementary excitations are magnons. However the magnons are strongly interacting and not always well defined: in addition to the one magnon branch, the two-magnon processes have important contributions to the dynamic spin structure factor (see Fig.\ref{Fig:BLBQ-chain}) \cite{Schmitt98}. Returning back to the critical point, the two-magnon excitations merge with the one magnon excitations and only a continuum of spinon excitations remain\cite{Schmitt98}.

Presumably, the qualitative features of the one dimensional model would carry over to a coupled two dimensional chain model. This implies that an AKLT chain state would similarly consist of both a continuum of excitations and a one magnon branch similar to those found in Fig. \ref{Fig:BLBQ-chain}. But the nematic quantum spin liquid in the main manuscript has no such one magnon branch: the spinons at the mean field level are deconfined. Distinguishing between the AKLT state and a nematic quantum spin liquid state is therefore a matter of finding evidence for the one magnon branch of excitations in neutron scattering. If such a signature exists, it will provide strong evidence for the AKLT chain state.

Finally, we should mention that the presence of itinerant fermions likely complicates this story as mentioned in the main text. The survival of the one magnon branch and or even the fundamental distinction between confined and deconfined spinons may disappear though arguments in the literature suggest an FL* state which preserves the fundamental distinction is possible\cite{Senthil03, Senthil04,Sachdev16}. 

\end{document}